\documentclass{emulateapj}
\usepackage{graphicx}
\usepackage{natbib}
\usepackage{graphicx}

\newcommand{\beq}	{\begin{equation}}
\newcommand{\eeq}	{\end{equation}}
\newcommand{\beqa}{\begin{eqnarray}}
\newcommand{\eeqa}{\end{eqnarray}}

\def\simlt{\lower.5ex\hbox{$\; \buildrel < \over \sim \;$}}
\def\simgt{\lower.5ex\hbox{$\; \buildrel > \over \sim \;$}}

\font\tenbi=cmmib10 
\newfam\bifam  \textfont\bifam=\tenbi

\font\tenbr=cmbx10
\newfam\brfam  \textfont\brfam=\tenbr

\font\squinttenbi=cmbx10 at 9pt
\scriptfont\brfam=\squinttenbi

\def\vecnabla{
              \setbox1=\hbox{$\bigtriangledown$}
                           \raise.45ex\hbox{$\bigtriangledown$\hskip-.97\wd1
                           $\bigtriangledown$\hskip-.97\wd1
                           $\bigtriangledown$\hskip-.97\wd1}
                           \raise.47ex\hbox{$\bigtriangledown$}}

\def\rsun{\ifmmode {\rm R}_{\mathord\odot}\else $R_{\mathord\odot}$\fi}
\def\msun{\ifmmode {\rm M}_{\mathord\odot}\else $M_{\mathord\odot}$\fi}
\def\lsun{\ifmmode {\rm L}_{\mathord\odot}\else $L_{\mathord\odot}$\fi}

\newcommand{\kms}	{{\rm km}\, {\rm s}^{-1}}

\def\tmb{\ifmmode {T_{\rm mb}^{13}(x,y,v)}\else $T_{\rm mb}^{13}(x,y,v)$\fi}

\shorttitle{Synthetic Observations}
\shortauthors{Offner et al.}
\begin{document}

\title{Observing Simulated Protostars with Outflows: How Accurate are Protostellar
  Properties Inferred from SEDs?}

\author{Stella S. R. Offner}
\affil{Harvard-Smithsonian Center for Astrophysics,
    Cambridge, MA 02138, USA}
\email{soffner@cfa.harvard.edu }

\author{Thomas P. Robitaille}
\affil{Max-Planck-Institute for Astronomy, K\"{o}nigstuhl 17, 69117 Heidelberg, Germany}

\author{Charles E. Hansen}
\affil{Astronomy Department, UC Berkeley, Berkeley, CA 94720, USA}

\author{Christopher F. McKee}
\affil{Physics and Astronomy Departments, UC Berkeley, Berkeley, CA
  94720, USA}

\author{Richard I. Klein}
\affil{Astronomy Department, UC Berkeley, Berkeley, CA 94720; Lawrence Livermore National Laboratory, Livermore, CA 94550, USA}

\begin{abstract}
The properties of unresolved protostars and their local environment are frequently
inferred from spectral energy distributions (SEDs) using
radiative transfer modeling. In this paper, we use synthetic
observations of realistic star formation simulations to evaluate the accuracy
of properties inferred from fitting model SEDs to observations. 
We use ORION, an adaptive mesh refinement (AMR) three-dimensional
gravito-radiation-hydrodynamics code, to simulate low-mass star
formation in a turbulent molecular cloud including the effects of protostellar
outflows. To obtain the dust temperature distribution and SEDs of the forming protostars, we post-process
the simulations using HYPERION, a state-of-the-art Monte-Carlo
radiative transfer code. We find that the ORION and HYPERION dust
temperatures 
typically agree within a factor of two. 
We compare synthetic SEDs of embedded protostars for a
range of evolutionary times, simulation resolutions, aperture sizes,
and viewing angles. 
We demonstrate that complex, asymmetric gas morphology  leads to a
variety of classifications for individual objects as a function of viewing angle. We derive best-fit 
source parameters for each SED through comparison with a
pre-computed grid of radiative transfer models.
While the SED models correctly identify the evolutionary stage of the
synthetic sources as embedded protostars, we show that the disk and stellar
parameters can be very discrepant from the simulated values, which is
expected since the disk and central source are obscured by the protostellar envelope.
Parameters such as the stellar accretion rate, stellar mass, and disk mass
show better agreement, but can still deviate significantly,
and the agreement may in some cases be artificially good due
  to the limited range of parameters in the set of model SEDs. Lack of correlation between the  
model and simulation properties in many individual instances cautions
against over-interpreting properties inferred from SEDs for 
unresolved protostellar sources.
\end{abstract}

\section{Introduction}


\begin{figure*}
\epsscale{1.19}
\plotone{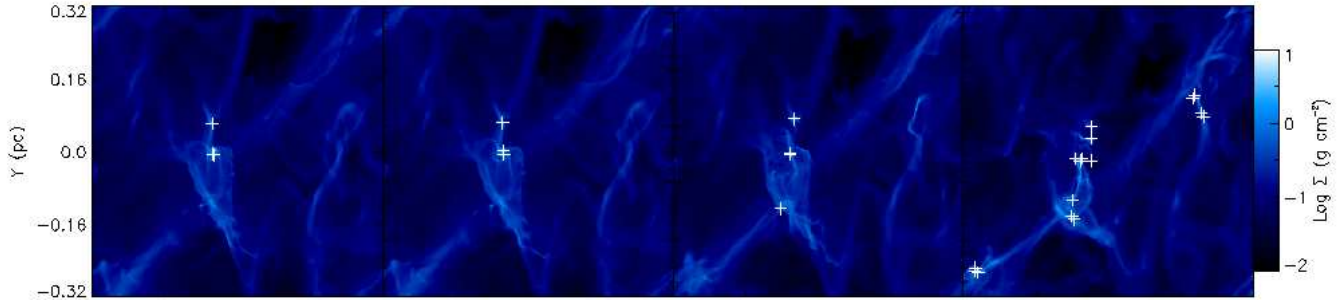}
\vspace{-0.85 cm}
\caption{Log column density for the 0, 15, 30 and 60 kyr zooms 
  from left to right. Star locations are
  indicated by the white crosses.
\label{column} }
\end{figure*}

Young protostars are deeply embedded in molecular gas and dust,
which leads to significant reprocessing of their radiation. Since most of these
sources cannot be directly imaged, it is necessary to infer properties of
the protostar, accretion disk, and gas envelope using other means. The proliferation of
infrared and millimeter instruments, such as Spitzer, 2MASS, CARMA,
Plateau de Bure, and
Herschel, has contributed a wealth of
multiwavelength data for numerous embedded protostars over the last decade. Spectral energy
distributions (SEDs), generally spanning $\sim$1-1000 $\mu$m, can be constructed
from this multiwavelength photometry. However, even with a
reasonably complete SED, complications such as excess
extinction, parameter degeneracies, and fitting subtleties elevate the
derivation of source attributes to an art form. 

SEDs are typically divided into four classes using either the
infrared spectral slope \citep{greene94} or the effective source (``bolometric'') temperature
\citep{myers93}. These classes generally serve as a proxy for the
  amount of radiation reprocessing that occurs due to the 
  embedding dense gas. 
Regardless of the details of the classification scheme, objects
characterized as Class 0 protostars are heavily obscured by dusty envelopes such
that most of the radiation falls in the sub-millimeter. Class I protostars are less obscured and may be surrounded by a massive 
circumstellar accretion disk. For a Class II object, some of the
direct stellar radiation is visible, and often the source is 
considered to be a pre-main-sequence star, which has accreted or expelled most of its initial
envelope and thus produces little (sub)millimeter emission. The
remaining gas lies in a thin circumstellar accretion disk. 
Outflow activity may affect the SED shape most significantly during the Class I
and Class II phases. For Class III objects, the disk dissipates and the
source approaches the main sequence.

While SED classification may be straightforward, deriving actual physical
properties from the SEDs is more complicated. For reasons discussed below, it is
very difficult to assign more than a rough evolutionary state to a given SED.
The most common means of inferring
properties involves picking a simple analytic model for the protostellar properties (e.g.,
mass, radius, and accretion rate) and
embedding gas morphology (e.g., disk mass, envelope mass, radial
profile, outflow cavity width) and then computing SEDs with
a radiative transfer code. The studies published by \citet[][henceforth R06]{robit06} and \citet{robitaille07}, greatly streamlined
parameter derivation by providing an extensive library of 200,000
aperture-dependent model SEDs defined by 14 physical parameters
(including a wide range of stellar masses and evolutionary stages) and
10 viewing angles for each physical model. The studies also included an
efficient way to fit these models to observations and to derive ranges of parameter values providing a good fit.
These models have been widely used on
problems ranging from estimations of individual
stellar properties \citep{robitaille07} to the Milky Way star formation rate
\citep{robitaille10}. 

Despite the breadth of parameter
space, the R06 models have limits. The morphologies are 
axisymmetric and thus do not account for gas turbulence, filamentary
structure, or outflow asymmetry. The stellar age is taken as one
parameter input, but the models do not form an evolutionary
sequence. (Indeed, the detailed evolutionary progression of an individual
protostar, including its formation time \citep{mckee10}, accretion
history \citep{Offner11a} and their subsequent effects on stellar
evolution \citep{hosokawa11} remains debated!).
Because of the inherent observational uncertainty in age estimations and parameter degeneracies, SED classes cannot be confidently mapped to stages of evolution (e.g., \citealt{crapsi08}).
The R06 models also
neglect gas with temperatures below 30 K, which may affect the fits at
longer wavelengths. Finally, the models assume that the source is a
single protostar. Although most low-mass stars in the field are single
\citep{lada06}, evidence suggest that protostars may have higher
multiplicity fractions \citep{duchene07}. In addition, estimating
protostellar multiplicity is challenging
even using high-resolution interferometry of local star forming
regions \citep{Offner11c}. In regions more than a kpc distant,
where each source is potentially a cluster of young protostars, the
multiplicity is impossible to obtain accurately (e.g., \citealt{kang09}).


There are also a number of complications to parameter derivations
that are independent of underlying model assumptions.
First, there is sizable parameter
degeneracy. This can be taken into account by considering the group of
models with reasonable fits, i.e., ``good-fit'' models,  rather than any single best-fit model.
Second, the source viewing angle has a significant effect
on the source classification. Even including an outflow cavity inclination
variable, it is very
difficult to distinguish an older edge-on Class II source from a
younger Class I source or a Class I source viewed along the outflow
axis from an older, less embedded Class II source. This type of
confusion can be reduced by making mass envelope estimates from
observations of the 
millimeter continuum \citep{enoch09} or by detecting molecular emission
associated with dense gas (e.g., \citealt{heiderman10, kempen09}), however, this data
is not always available.

The goal of this work is to model the SEDs of sources in realistic,
non-magnetized simulations and explore the challenges of parameter
estimation from SEDs based on idealized models (e.g., R06).
We use ORION, an adaptive mesh refinement (AMR) three-dimensional
gravito-radiation-hydrodynamics code, to simulate low-mass star
formation in a turbulent, clustered environment. The simulations
include both the effects of radiation feedback from the forming stars and protostellar
outflows.  In order to obtain model
dust temperatures and synthetic SEDs for the sources, we post-process the simulations using the HYPERION
radiative transfer code \citep{robitaille11}. Section 2 contains an
overview of our numerical methods. In section 3 we present the SEDs
for four evolutionary times and a variety of simulation resolutions,
viewing angles and aperture sizes. We also derive best-fit 
source parameters for each SED through comparison with the R06
grid of models and then compare these with the actual simulation values. We conclude in section 5.

\begin{figure*}
\epsscale{1.17}
\plotone{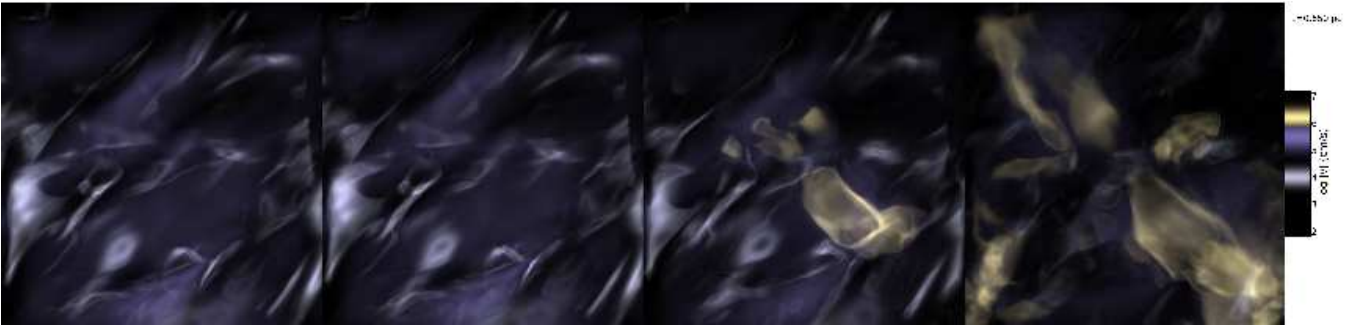}
\caption{Volume rendering of gas velocity for the 0, 15, 30 and 60 kyr zooms 
  from left to right.
\label{windvel} }
\end{figure*}

\section{Methods}

\subsection{3D Radiation-Hydrodynamic Simulations}

We perform self-gravitating radiation-hydrodynamic calculations
using the ORION Adaptive Mesh
Refinement (AMR) code \citep{truelove98, klein99}. Our simulations adopt the same initial
conditions as \citet{Offner09} and numerical procedure as
\citet{hansen12}, which we briefly summarize below. 

The simulations are of a turbulent box with a mean density of $4.46\times10^{-20}$
g cm$^{-3}$, a total gas mass of 185 $\msun$, cloud length of 0.65 pc,
and an initial 3D Mach number of
6.6.  
To initialize the domain, we inject energy at a constant rate for
two shock crossing times without
self-gravity (e.g., following \citealt{stone98}). After this turbulent driving phase, gravity is turned on and
the turbulence is allowed to naturally decay.  

Sink particles are inserted in regions of the
flow that exceed the Jeans condition \citep{krumholz04}. 
These ``stars'' are endowed with a sub-grid stellar evolution
model that includes accretion luminosity down to the stellar
surface, Kelvin-Helmholz contraction, and nuclear burning. 
A second sub-grid
model based upon \citet{matzner99} governs the launching of protostellar winds. 
Our sub-grid models  
allows us to include all radiative heating in a way that is
independent of the simulation resolution. Since we also include the
effects of  protostellar
outflows, the accretion rate from the disk onto the star
is reduced from the case without outflows, and the accretion luminosity is thus smaller than computed in
\citet{Offner09} by a factor of
$\sim 10$ \citep{hansen12}.
As pointed out by \citet{bate11}, one defect of stellar sub-grid models
is that the luminosity depends explicitly on the protostellar radius. This
radius can vary by a factor of two depending upon the details of the
adopted 
evolutionary model, including the  assumed
radiative efficiency, initial seed radius or mass, and gas
opacities. Nonetheless, this uncertainty corresponds to a factor of only 
$2^{1/4}$ ($<20$\%) in gas temperature \citep{Offner09}.

We summarize the key aspects of our outflow model below but refer the reader to
\citet{cunningham11} for the full details of the model implementation in ORION.
In our outflow model, the wind launching velocity is given by the Keplerian
velocity at the stellar surface, $v_K = \sqrt{G M_*/r_*}$. For 
high-mass protostars these velocities can exceed $200~\kms$, greatly
limiting the numerical time step of the
calculation. \citet{cunningham11} circumvent this issue by capping the
outflow velocity at a fixed fraction of the Keplerian speed. 
In the calculations we present here,  we instead economize the
computational expense by capping the launching velocity, $v_K \le$60 km s$^{-1}$, and
only proceeding to high resolution at discrete times.
 
Note that in our wind model the outflow momentum and direction are determined
by the instantaneous protostellar mass, accretion rate, and angular
momentum vector. Once deposited on the grid, the gas
evolves hydrodynamically, which leads to outflow morphology and
asymmetry similar to that of observed outflows \citep{Offner11}. 

For the dust opacities, we adopt a dust model that assumes
a standard iron abundance and treats the grains as composite
aggregates \citep{semenov03}. This model produces an extinction ratio of $R_v=3.42$ for dust with
temperature $T<120$K. In most of the domain, dust serves as the
dominant coolant. However, strong shocks produced by outflowing gas running into ambient material
can result in cells with temperatures exceeding the dust
destruction temperature of $\sim 1000$K. In this regime, we
implicitly calculate the gas temperatures assuming atomic line cooling
\citep{cunningham11}.

We perform the numerical calculations as follows.
First, we run a simulation of a forming cluster for one freefall time with a  maximum cell resolution of
128 AU (two AMR levels). We then ``zoom'' in on four different time
slices by restarting the calculation and allowing it to evolve with 
five additional AMR levels (4 AU resolution). This is similar to the procedure in
\citet{Offner11} except that here all protostars receive extra
resolution. 
New grid cells are added to satisfy a Jeans condition with $J=0.125$ \citep{truelove97} and a geometric refinement
criterion requiring that each star is centered within a block of 8$^3$ fine cells. 
We refer to the high resolution calculations according to the time
when the zoom begins relative to the formation of the first
star in the simulation: 0 kyr, 15 kyr, 30 kyr and 60 kyr. Extra resolution is added to the first time slice beginning just prior to
the formation of the first star particle. During each zoom, we evolve the
calculation for $\sim$ 8 kyr, where we allow structure to develop in the
newest cells before proceeding to the next AMR level. Table
\ref{stars} gives the protostellar masses and luminosities at the
highest resolution during each of the zooms. In our analysis,
the oldest star is $\sim70$ kyr, so we consider only the earliest
embedded stage of star formation in this work.

Figure \ref{column} shows the column density at the completion of
each zoom. At the final time, there are 20 stars, which are clustered
in several different regions of the domain.
Figure \ref{windvel} shows a volume rendering of the gas velocity
for each of the zooms. As the calculation evolves, the outflows of
the first forming stars extend across the domain and dominate
the turbulent motions. By the final time, the outflows are too entangled to
separate individually, and the simulation resembles an outflow
dominated cluster like NGC1333 \citep{swift08}.

\begin{deluxetable}{lccc}
\tablewidth{0pt}
\tablecolumns{3}
\tablecaption{Star particle properties at the highest resolution for
  each of the zooms. The stars are listed in order of formation.\label{stars}}
\tablehead{Time (kyr) & Mass $(\msun)$ & Luminosity $(\lsun)$ &
  Name\tablenotemark{a}}
\startdata
0 & 0.032  & 0.11 & 1a\\
  & 2.0e-6 & - &\\
  & 7.8e-5 & -& \\
\hline
15 & 0.053  & 0.25& 1b\\
 & 0.023  & 1.06 & 2a \\
 & 1.2e-5 & - &  \\
 & 0.012 & 0.23 & 3a\\
\hline
30 &0.19 & 3.19 &   \\
   &0.14 & 1.50 & 2b  \\
   &0.64 & 1.86 & 1c \\
   &0.05 & 0.23 & 3b \\
\hline
60 & 0.73  & 77.71\tablenotemark{b}& \\
& 0.29 & 0.35 &  2c\\
& 1.14 & 10.66 & 1d\\
& 0.12 & 1.11 & 3c \\
& 0.13 & 0.26 & 4 \\
& 0.11 & 0.36 & 5 \\
& 0.17 & 1.059 & \\
& 0.065 & 0.24 & 6\\
& 0.058 & 0.13 & 7\\
& 0.10 & 1.15 & 8\\
& 0.52 & 7.39 & 9\\
& 0.078 & 0.13 & \\
& 0.077 & 0.18 & 10\\
& 0.047 & 0.20 & 11\\
& 0.19 & 1.86 & 12\\
& 0.10 & 0.21 & \\
& 0.038 & 0.11 & 13\\
& 0.037 & 0.21 & 14\\
& 0.052 & 0.25 & 15\\
& 0.036  & 0.13 & 16\\
\enddata
\tablenotetext{a}{See Figure 6. Only those with sufficient luminosity and those that do not have a nearby, more massive companion are assigned a unique name.}
\tablenotetext{b}{This source is brighter than its more massive
  companion (1d) because its instanteous accretion rate is higher.}
\end{deluxetable}

\subsection{Spectral Energy Distribution Modeling}

We use HYPERION, a parallel 3D Monte-Carlo dust continuum radiative transfer code, to compute synthetic
spectral energy distributions for the forming stars in the simulations 
(see \citealt{robitaille11} for algorithmic details). Since the HYPERION infrastructure is designed to be as
generic as possible, we are able to map the ORION AMR cells to
a tree structure that can be directly read by HYPERION without interpolation
or information loss. Moreover, the sources do not need to be excised
from the domain and computed separately. For each time slice, HYPERION
calculates the SED for a set of apertures centered on each specified
source as viewed through the box. In some cases, especially for the
larger apertures, multiple sources may contribute to the net emission.
The HYPERION inputs are the dust density as a function
of position, an opacity model, source luminosities and source
temperatures. The dust temperature is calculated by HYPERION
based upon the input sources. For consistency, we use the same
aggregate 
dust grain opacity model used in the ORION simulations and adopt the
coefficients calculated for dust with temperature $<120$K
\citep{semenov03}. 

When performing the radiative transfer calculation, we exclude any star particles that have
not yet begun burning Deuterium. These star particles tend to have such small masses that
their accretion luminosity is negligible. (These stars are shown in
Figure \ref{column} for completeness.)
We also zero the density within 8$^3$ fine cells immediately
around each star. This volume contains the accretion and
outflow launching regions. Numerical effects reduce the accuracy of
these cells, so that they may be over-dense above and below the disk plane, where the wind
would otherwise sweep out a cavity. This is discussed in more detail
in section \ref{innerdisk} and Appendix A.

We perform the radiative transfer calculations with 10$^7$ photon
packets to ensure that the shorter wavelength fluxes are reasonably
well converged. For this number, additional photon packets do not significantly
change the SED. The SEDs are
computed with 200 wavelengths logarithmically distributed between 0.1
$\mu$m and 5000 $\mu$m. For each
source and viewing angle, we compute the flux for 20 circular
apertures with radii logarithmically spaced between 1000 AU and
20,000 AU. Each source is viewed from 20 perspectives, which are
regularly spaced in 
$\theta$ and $\phi$ \footnote{Geometrically, 20 regular angular spacings
  in spherical
  coordinates correspond to the vertexes of a dodecahedron.}.

\section{Synthetic Observations}

\begin{figure*}
\epsscale{1.15}
\plotone{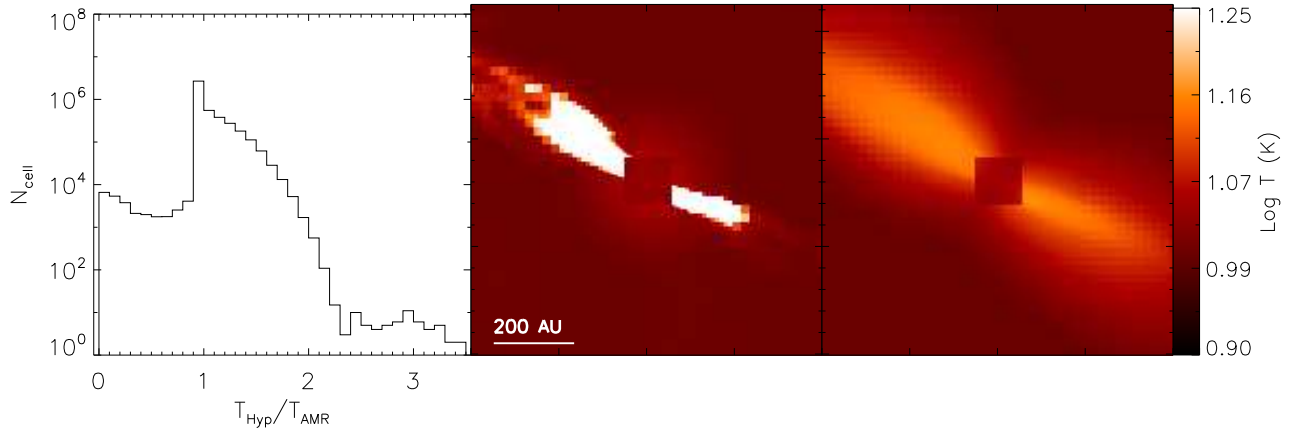}
\caption{Estimated dust temperature for the first forming protostar at
  0 kyr (just after the formation of the first sink particle). Left: Histogram of the temperature ratios for
  all cells in a (1400 AU)$^3$ region centered on the protostar. 
Center: Temperature slice showing the ORION dust temperatures. Right: Temperature slice showing the HYPERION dust
  temperatures. The central $16^3$ finest cells have been excised (this is where the
  HYPERION density input was set to zero).
\label{temp} }
\end{figure*}

\subsection{Dust Temperature Comparison}

HYPERION calculates the dust temperature, $T$, by assuming local
thermodynamic equilibrium (LTE), wherein the dust emission balances
the specific energy absorption rate, $\dot A$:
\beq
4 \pi \kappa_{\rm P}(T)B(T) = \dot A, \label{lte}
\eeq
where $\kappa_{\rm P}(T)$ is the pre-computed Planck mean mass absorption
coefficient and $B(T)$ is the integrated Planck function. The right
side, which depends upon the mean intensity, $J_{\nu}$, can be
expressed as a function of the energy per photon packet, $\epsilon$,
cell volume, $V$, emission time, $\Delta t$, photon path length
between events, $l$, 
and mass absorption coefficient \citep{lucy99}:
\beq
\dot A = 4\pi \int_0^{\infty} \kappa_{\nu}J_{\nu}d \nu = {1\over{\Delta t}} {\epsilon \over V} \sum l \kappa_{\nu}.
\eeq
On top of this solution, we impose a minimum dust temperature of 10 K.
For HYPERION, stars are the only heating source. Dynamical
effects such as shock and viscous heating are
neglected by eq.~\ref{lte}. Viscous heating is much smaller than radiative
heating, where any significant contribution falls mainly within the
disks 
\citep{Offner09}. However, shocks may significantly heat the gas,
particularly affecting outflow gas more distant from the star.

In comparison, ORION makes several different assumptions in calculating the gas
temperature. First, it assumes that the dust and gas are
perfectly collisionally coupled and, hence,  have the same temperature
and velocity. ORION assumes that dust is the
primary coolant (except in superheated
regions in which the dust has sublimated). To obtain the temperature, ORION solves the non-equilibrium, flux-limited,
radiation diffusion equation:
\beq
{{\partial E}\over{\partial t}} - \nabla \cdot \left({{c
  \lambda}\over{\kappa_{\rm R} \rho}} \nabla E \right) = \kappa_{\rm R} \rho
\left[ 4 \pi B(T)-cE \right]
+ \sum_n S_n,
\eeq
where $E$ is the radiation energy density, $\kappa_{\rm R}$ and
$\kappa_{\rm P}$
are the Rosseland and Planck mean dust opacities, $\lambda$ is the
flux-limiter, 
$\rho$ is the gas
density, and $S_n$ are the stellar
sources. This formulation of the diffusion equation omits velocity
dependent terms that account for the advection of radiative enthalpy,
since these are not significant for low-mass stars. 
At each timestep, the equation is solved iteratively until
the temperature converges in each cell. The density is determined
hydrodynamically and remains constant during the iterations. The
opacities are time-lagged and also remain constant. We impose Marshak boundary
conditions at the edge of the domain, which corresponds to a 10 K
radiative flux. This ensures that the gas temperature limits to 10 K
far from the stellar sources.
Radiative diffusion is generally a good approximation within
dense cores where the gas is fairly optically thick. 


Figure \ref{temp} shows a comparison of the estimated dust
temperatures around a young protostar. In both cases, the heating is
mainly confined to the outflow cavity. In HYPERION, the difference
between the core dust and outflow dust is due to radiative beaming. In
ORION, although some beaming can occur (e.g., \citealt{cunningham11}),
the elevated outflow temperature is due
primarily to shocks. Consequently, these cells are a factor of $\sim2$
warmer than estimated by HYPERION. Similar differences are apparent at
later times.

\subsection{The Spectral Energy Distribution Zoo}

In this section we examine the evolution of the source SEDs as a function of time, resolution, aperture, and viewing
angle. 

\subsubsection{Resolution}

\begin{figure}
\epsscale{1.15}
\plotone{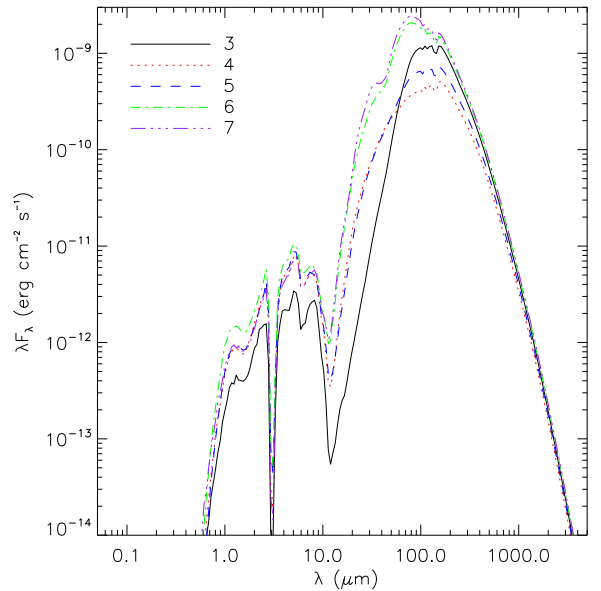}
\caption{A 30 kyr source with SEDs calculated at fixed inclination and 1000 AU
  aperture. The curves range from 
  low (maxlevel = 3, 65 AU) to high (maxlevel=7, 4AU)
  resolution. Consecutive 
  SEDs are $\sim$ 1.5 kyr apart.
\label{zoom} }
\end{figure}

Figure \ref{zoom} shows a source observed at each zooming stage. To
allow structure to develop at each resolution, there is an offset of
$\sim 2000$ kyr between the zoom images.\footnote{New fine cells are
  initialized by interpolating data from the parent lower resolution cells. Without
  additional hydrodynamic evolution, the data on the finest level cannot be
  considered higher resolution.} For consistency, we impose an inner
gas radius of 16 AU (4 cells at the finest resolution zoom) for all zooms. At
higher resolution, the main emission peak in the far-infrared shifts to shorter
wavelengths. Progressively more emission escapes in the near-infrared
as the central disk becomes more compact and the cavity extends
closer to the star (see Section 3.3). However, the trends aren't
entirely monotonic with resolution. At long wavelengths the SED is
fully converged and emission comes from larger scales. Consequently, changes in
the sub-millimeter portion of the SED partially result from the time evolution
between each resolution snapshot. The effects of time evolution and
additional resolution cannot be disentangled.

\subsubsection{Aperture}

\begin{figure}
\epsscale{1.15}
\plotone{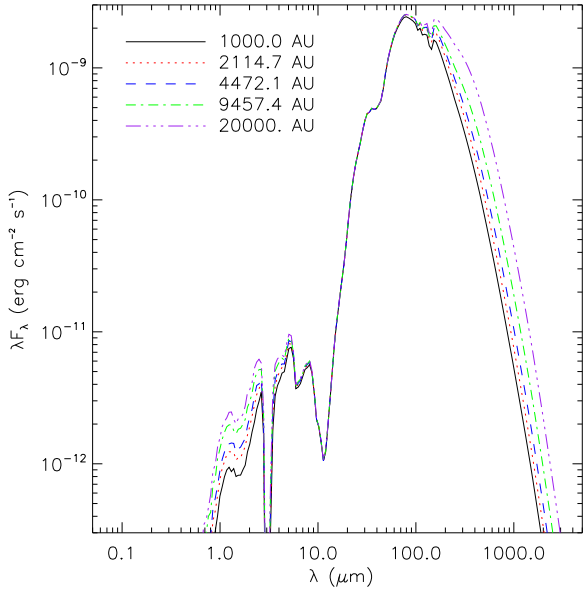}
\caption{ Source at 30 kyr with SEDs at fixed inclination (same as in
  Figure \ref{zoom}) and at the maximum resolution viewed with five
  different aperture sizes. 
\label{aps} }
\end{figure}

The choice of aperture can be observationally useful for probing
emission at specific unresolved size scales.  Since the long
wavelength emission on larger scales primarily derives from the core envelope,
emission differences between
large and small apertures can be used to estimate the envelope
gas mass. 
For multiple apertures, an average density profile can also be
inferred by fitting how quickly the emission decreases with
aperture size.  Figure \ref{aps} illustrates the monotonically
shrinking emission at long wavelengths as the aperture size decreases. 
The near-infrared emission also decreases slightly. This is due to the
radiation emitted closer to the star scattering to larger scales.

\begin{figure*}
\epsscale{1.125}
\plotone{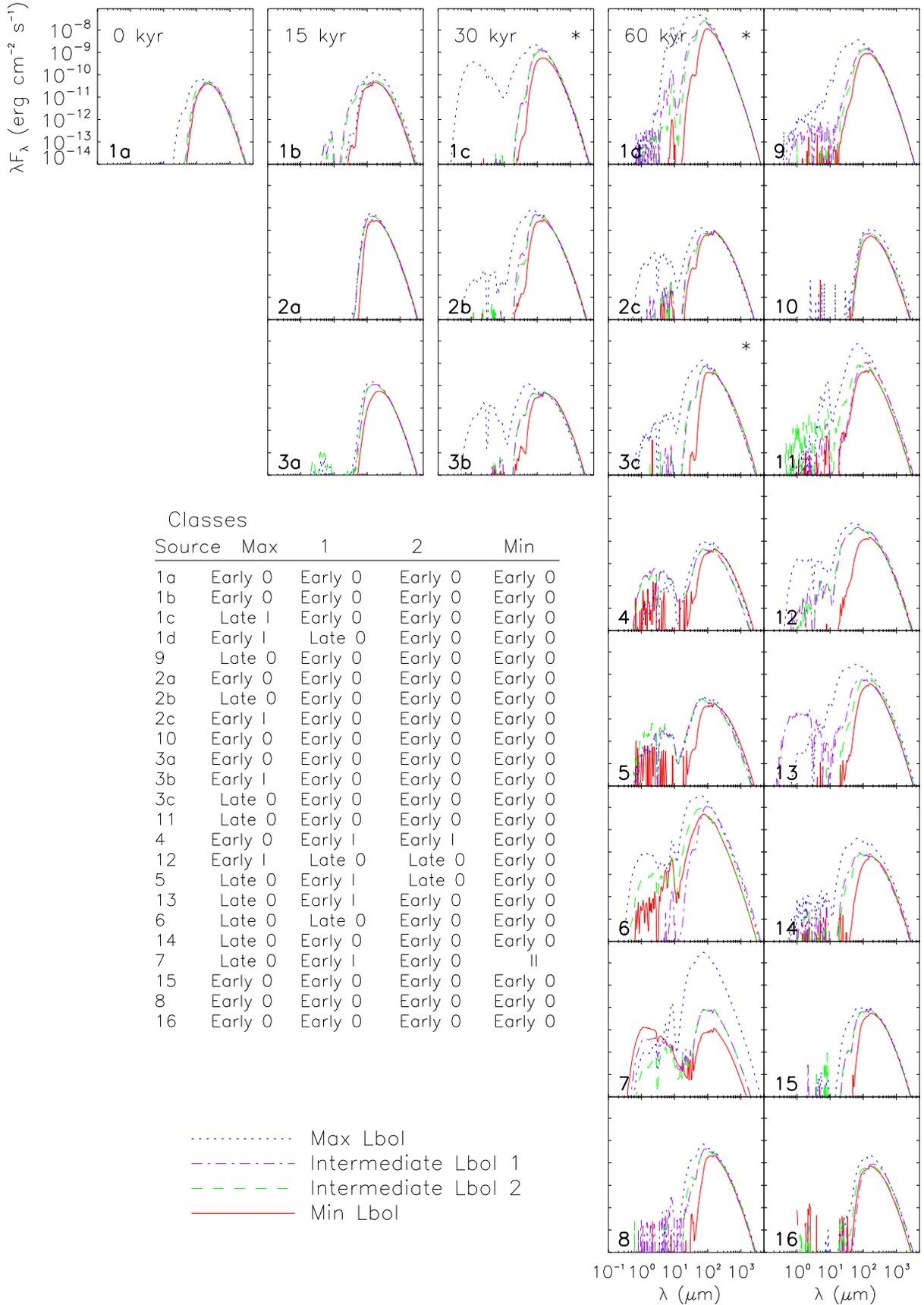}
\caption{Sources at the four different times viewed through a 1000 AU
  radius aperture. The output time for each column is denoted in the
  first row. Each panel contains SEDs for four different viewing
  angles: the SEDs corresponding to the minimum and maximum
  bolometric luminosities and two intermediate bolometric
  luminosities. If there is a secondary protostar within 1000
  AU only the SEDs for the primary source are depicted; these panels are marked
  with an asterisk. The inset table indicates the class of each SED.
\label{seds} }
\end{figure*}

\subsubsection{Time Evolution}

Star formation models suggest a basic progression in the source
properties over time: the envelope mass decreases, the outflow cavity
widens allowing more stellar radiation to escape, and the disk
shrinks. However, exactly how this progression occurs and specifically
how this impacts the SED in not well known.
The R06 analytic prototypes for the source, disk, and envelope
parameters encompass such changes broadly,
without specifying an exact evolutionary sequence. 
Our simulated protostars, however, do follow a consistently calculated
evolutionary progression. 
Asymmetric gas morphology  
and non-monotonic evolution, which results from effects such
as variable stellar accretion, add another level of complication over
the R06 models. In this section we consider the time evolution
of the SEDs in our simulations and the effect on underlying properties.

Figure \ref{seds} shows the SEDs of the forming stars at four different
times. The first forming star appears at all four times, while the
second and third star appear at three times.
In a few cases, there is a secondary or tertiary star present within the
observing aperture. In these cases, the SEDs of the two sources are nearly identical, so
we present only the SED for the most massive.
Each panel shows the seds for the source viewed at four different inclinations (see
section \ref{inclination} for discussion). These SEDs correspond to the
views with the
highest and lowest bolometric luminosities and to two views near the
median inferred bolometric luminosities.

Figure \ref{seds} illustrates several anticipated broad trends. The
spectral peak shifts to shorter wavelengths with time. This shift is
partially due to more direct and scattered emission escaping at
shorter wavelengths. More viewing angles sample this emission with
time (see section \ref{view}). Nearly all the sources at 60 kyr show
some emission at shorter wavelengths, even the youngest sources. This
is because star formation within the cloud is somewhat clustered, as
illustrated by Figure \ref{column}, and
the interacting outflows of the older sources impact the environment of
the younger sources (see Figure \ref{windvel}).

For the sources with close companions, we find that the SED appearance is not directly indicative of a second
nearby star. Without resolving the second source it is thus impossible to
tell which of the sources have companions from the SED shapes alone.
Disentangling the effects of multiplicity on the SED is an
important problem, but it is beyond the scope of our current study.

\begin{figure}
\epsscale{1.2}
\plotone{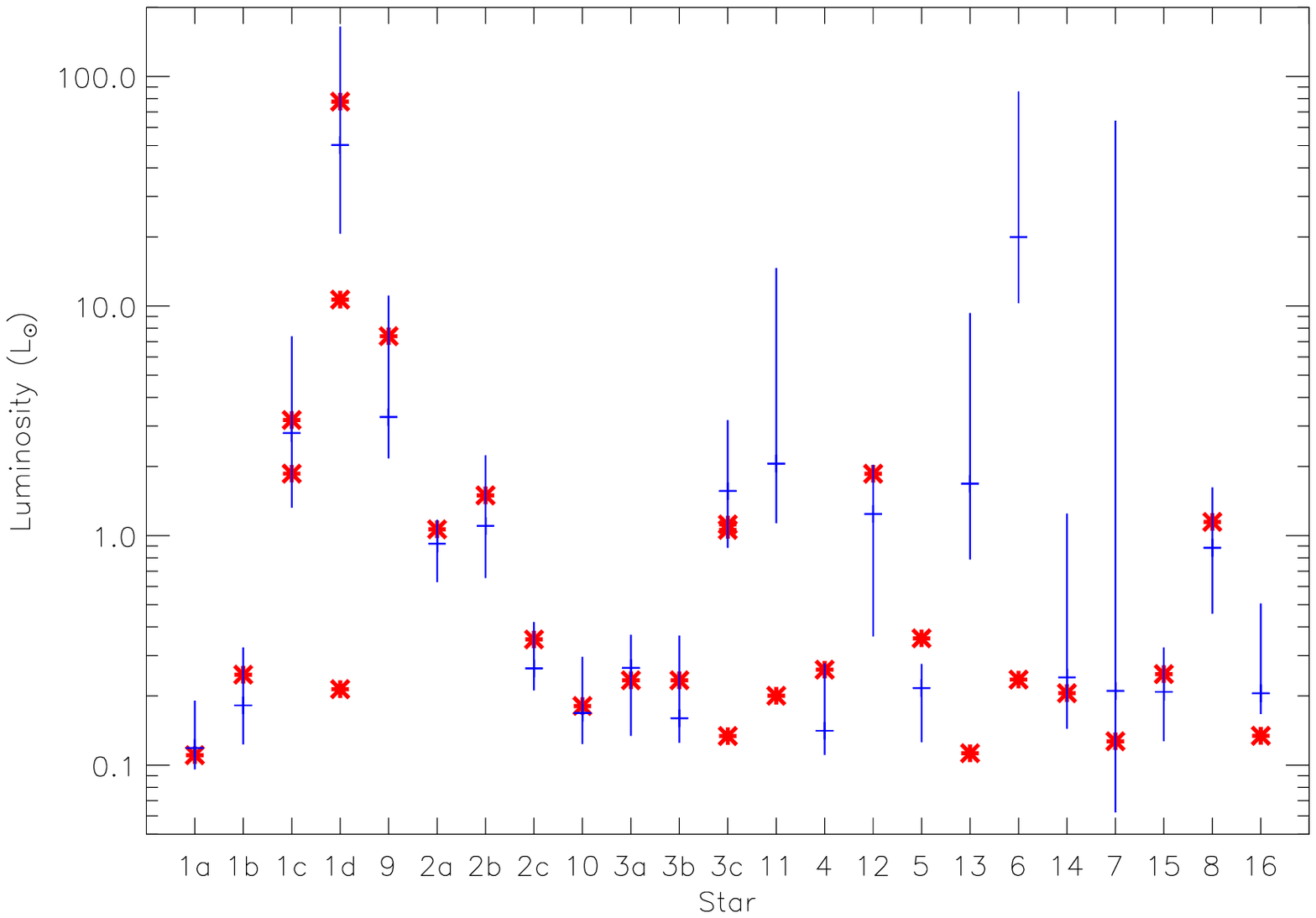}
\caption{Bolometric luminosities for all sources shown in Figure \ref{seds}. The input
  luminosity (simulation) for each model is indicated by the red
  stars. The vertical range indicates the  
  bolometric luminosities spanned by the different views,
  where the 1st intermediate bolometric luminosity is marked by the
  horizontal line. The aperture radius is 1000 AU. The luminosities
  of sources within 1000 AU of the primary are also shown (e.g., 1c,
  1d, 3c).
\label{lum_range} }
\end{figure}

Figure \ref{lum_range} shows the wavelength-integrated flux, i.e., luminosity, derived from the SEDs for
each source and the range between the minimum
and maximum. In some cases the range of
luminosities spans an order of magnitude. For source 7, it spans three
orders of magnitude. In almost all
cases, the input luminosity falls within the observationally estimated
range. In the three cases where there is large disagreement (sources
6, 11, 13), the observed protostars are near more luminous
sources. Although none of these is physically within 1000
AU of another source, the apertures may still overlap significantly with that of a brighter
source, increasing the net emission. For example, source 6
lies $\sim$ 1300 AU away from the brightest source, 1d. Since it is
relatively close to 1d and dim by comparison,
it is unlikely that it would be identified as a separate source observationally.

Increasing the aperture size has two effects on the luminosity
ranges shown in Figure \ref{lum_range}. First, as more envelope gas is included, the integrated luminosity
increases. For large apertures, i.e., $\ge 10,000$ AU, the minimum
bolometric luminosity can be 2-3 times the actual protostellar
luminosity (even accounting for the presence of a second
protostar). The excess emission arises from the core envelope, which
is radiating at $\sim$10 K (e.g., \citealt{enoch08}). 
Second, if a significant envelope remains, the IR flux
contributes less to the total luminosity for large apertures such that the difference between the
minimum and maximum bolometric luminosities decreases.

\subsubsection{Viewing Angle}\label{view}

Previous comprehensive SED modeling has highlighted the influence of
viewing angle on the 
inferred luminosity, SED shape, and apparent age (e.g.,
\citealt{whitney03a}). We find a similar dependence here. Figure \ref{seds}
illustrates SED variability with source
inclination. For example, one view observes source 1c down the
outflow cavity, making it appear much warmer and more evolved
than its actual age of $\sim35$ kyr. In most cases, however, there is
no clear sight line to the star since the outflow cavity is not entirely
free of dust and gas. As a result, we find slightly less excess IR
emission than idealized models for this view (e.g., \citealt{whitney03a}).

Three of the sources (numbers 1, 2 and 3) are present during more than one
zoom. The viewing angle corresponding to the SEDs with the minimum and maximum
bolometric luminosities shown in Figure \ref{seds} is not necessarily
the same across these panels.
Since the sources are not fixed with respect to the grid,
the orientation at which an observer would view the
source, for example down the outflow cavity, evolves with
time. However, the brightest and dimmest SEDs generally
correspond to the views down the cavity and through the edge-on disk, respectively. 

\begin{figure}
\epsscale{1.15}
\plotone{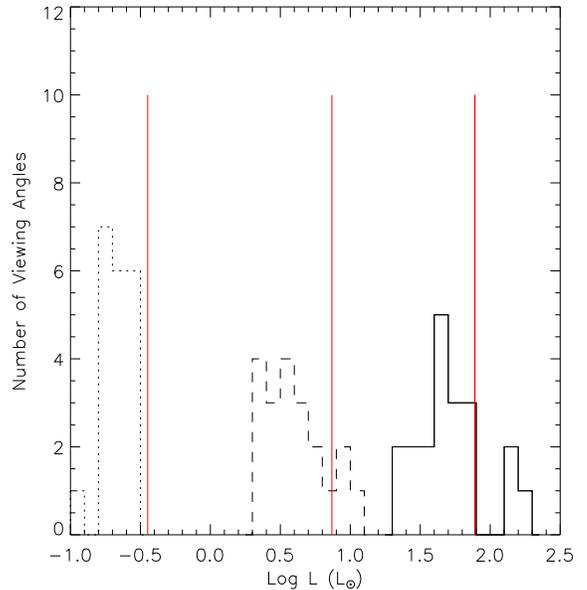}
\caption{Histogram of bolometric luminosities for the 20 different viewing
  angles for sources 4 (left), 9 (middle) and 1d (right). The input
  luminosity (simulation) for each model is indicated by the red vertical lines. The
  aperture size is 1000 AU. 
\label{lbolhist} }
\end{figure}

Figure \ref{lbolhist} shows some typical luminosity distributions for 
20 observed viewing angles. Younger more embedded sources tend to
have a lower luminosity dispersion. This trend is apparent for source
1 in Figure \ref{seds}. However, the picture becomes more confused when the sources
are clustered. The environment of the younger protostars is contaminated by 
outflows and heating from older nearby objects.

\begin{figure}\epsscale{1.2}
\plotone{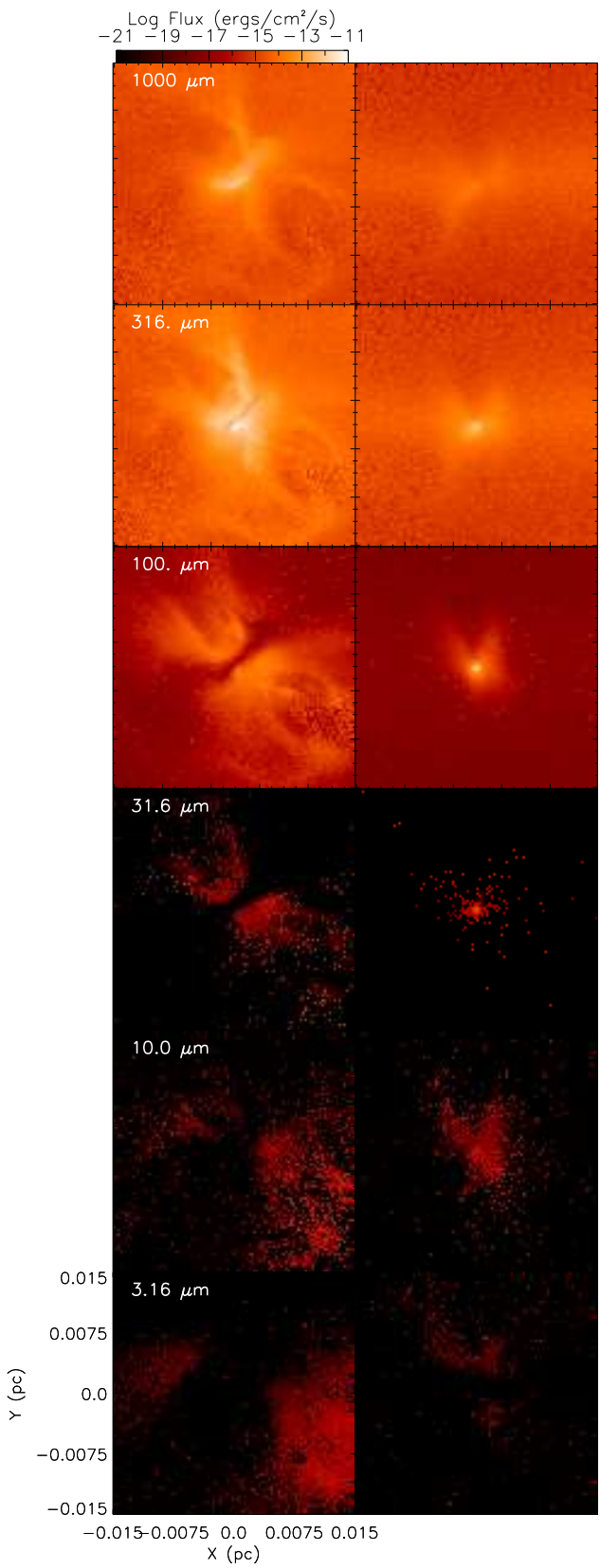}
\caption{Two protostars in the 30 kyr time slice imaged in six different wavebands.The first forming
  star in the simulation is on the left.
\label{images1} }
\end{figure}

\begin{figure}
\epsscale{1.2}
\plotone{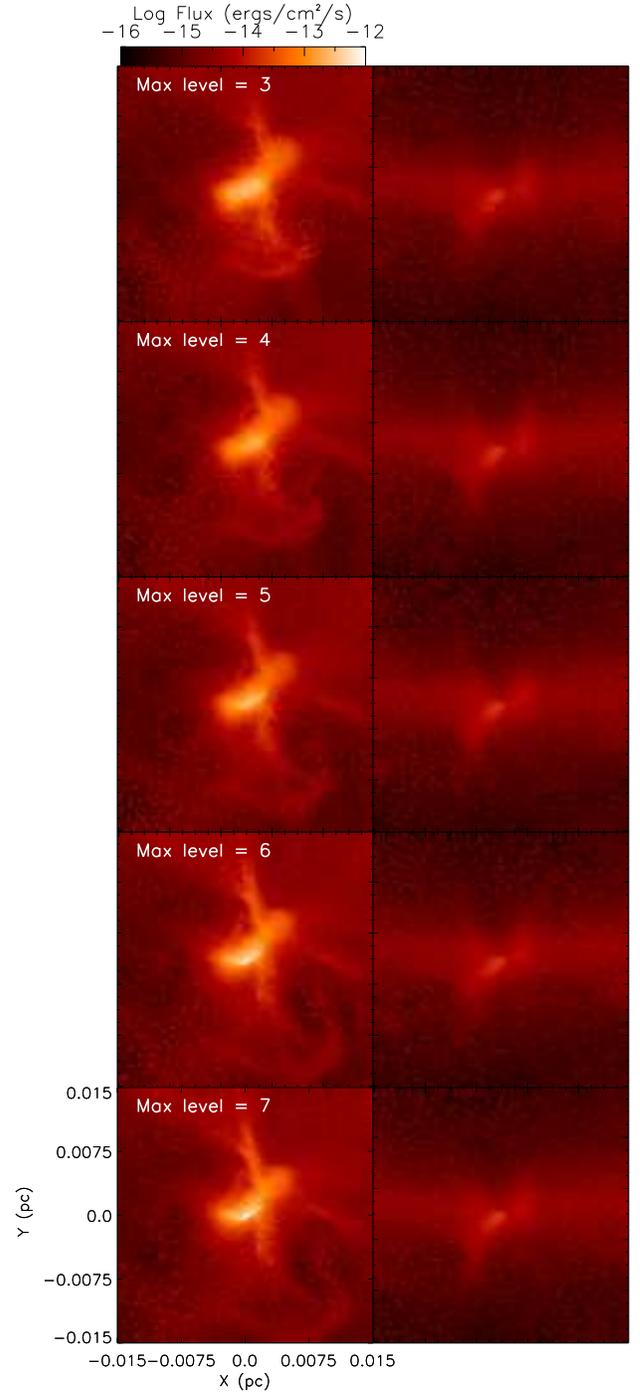}
\caption{Two protostars in the 30 kyr time slice (the first forming
  star is on the left) imaged in 1mm, where the AMR maximum resolution
  increases from 65 AU (top) to 4 AU. The images are $\sim$
  1.5 kyr apart.
\label{images2} }
\end{figure}

\subsection{Protostellar Imaging}

The origin of the emission contributing to the SED shape is apparent
in images of the constituent wave bands. Figure
\ref{images1} shows two protostars viewed in six different
wavelengths. The longer wavelengths pick out the colder envelope
dust. In the mid-infrared, the cooler, edge-on disk becomes visible
as a narrow extinction band perpendicular to the outflow.
Towards shorter, near-infrared wavelengths, scattered light highlights 
the outflow cavity. The radiative beaming caused by the outflow cavity 
is readily apparent. As sources become older and the dust warms, 
emission increases at shorter wavelengths. The images of the younger
source, which is not edge-on, indicate a much narrower outflow cavity.

Figure \ref{images2} illustrates the effect of increasing the
simulation resolution. Overall, the higher resolution has a small effect on the outflow morphology. However,
high density regions, such as the disk, become denser and more compact.

\subsection{Source Classification}

In this section, we compare the source evolutionary stage derived from observational metrics
with the actual source properties. There are two main, arguably comparable, ways to
assign a spectral Class to a source. The first involves computing the
spectral slope of the SED in the near to mid IR \citep{greene94}. Young sources have a
steeply increasing slopes ($\alpha \ge 0.3$) that flatten and then
decline ($\alpha < -1.6$) as the source evolves and
accretes its envelope. 
In this work, we use the characteristic spectral temperature, or ``bolometric temperature'',
to determine the spectral Class.

The bolometric temperature, $T_{\rm bol}$, is defined as
\begin{equation}
T_{\rm bol} = 1.25 \times 10^{-11} \langle \nu \rangle {~~\rm K},
\end{equation}
\citep{myers93} where the mean frequency is given by
\begin{equation}
\langle \nu \rangle= {{\int \nu S_{\nu}d\nu}\over{\int S_{\nu}d\nu}},
\end{equation}
where $S_{\nu}$ is the flux density.
Low bolometric temperatures correspond to young, dim protostars with
relatively massive, cold envelopes. The bolometric temperature gradually approaches
the effective stellar surface temperature as the surrounding envelope is accreted or expelled.
We separate the SEDs into early and late Class 0, early and late
Class I, and Class II (e.g., \citealt{enoch09}; see Table 2 for definitions). The exact division of bolometric
temperatures into classes is a matter of definition; in some cases 
$T_{\rm bol} =$70 K serves as the dividing line between Class 0 and
Class I objects \citep{enoch09, evans09}. Here we use 100 K as the
cutoff between late Class 0 and early Class I objects.

\begin{figure}
\epsscale{1.2}
\plotone{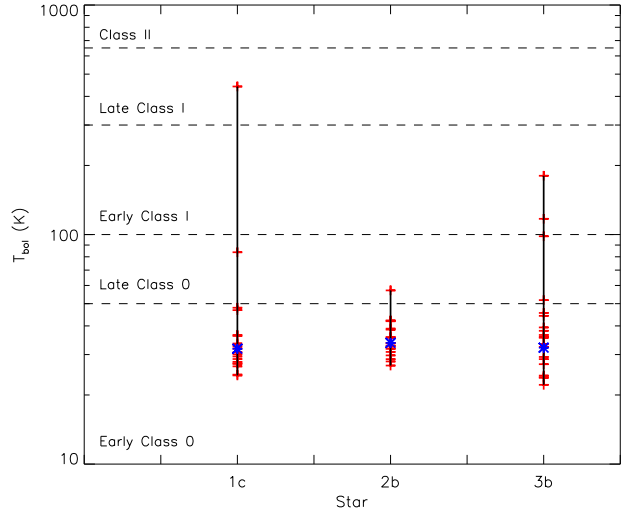}
\caption{Bolometric temperatures and corresponding classification
  for three sources at 30 kyr observed in a 1000 AU radius aperture. The vertical bar indicates the range of values for
  20 different views (red), while the blue star indicates the
  median value. 
\label{tbol} }
\end{figure}

Figure \ref{tbol} shows the inferred spectral Classes for several
sources. The oldest (1c) shows a larger range of inferred Classes,
including one view corresponding to late Class I. This indicates
that even if a source is $\lesssim$30 kyr old, which is below the estimated
Class 0 lifetime of 40-100 kyr \citep{enoch09, maury11}, it may appear as a Class I object $\sim$5-10\%
of the time. Statistically, some of these apparently older sources may
be offset by older sources obscured by their disks, which appear younger.
Reassuringly, the 
sources in Figure \ref{tbol} appear to be early Class 0 from most  
inclinations. All the sources have a minimum bolometric temperature
around 20 K, which is reasonable given that the minimum dust temperature is
10 K.

Table \ref{tabletbol} summarizes the number of sources observed in each
class at each time, assuming that every view corresponds to an
independent observation. In reality, none of these sources are older
than the typical
estimated Class 0 lifetime, so it is surprising how
many apparent Class I ($\sim$8\%) and Class II ($\sim$1\%) sources
occur. If the estimated Class 0 lifetime were $\sim$ 70 kyr or less and the
star formation rate were constant, then we would expect equal numbers
of early and late Class 0 sources when observing the cloud at the last
output time. We find that 70\% are early Class 0 sources, suggesting that
the Class 0 lifetime is longer 
(although this argument implicitly
assumes that the sources spend an equal amount of time in the early
and late stages, which may not be the case).

In order to diminish the role of viewing angle on the classification,
\citet{crapsi08} propose using a minimum envelope mass of $0.1\msun$ to
distinguish between Class I (younger) and Class II (older) sources, i.e.,
  between young embedded protostars (Stage I in the terminology of R06) and older disk-dominated
  sources (Stage II).  
We can estimate the envelope mass using 
\begin{equation}
M_{\rm env} = {{d^2 S_{\rm 1mm}}\over{B_{1mm}(T_{\rm D})\kappa_{\rm
      1mm}}}, \label{menv}
\end{equation}
where $d$ is the cloud distance, $S_{\rm 1mm}$ is the flux density at
1.1 mm observed with a 30'' diameter aperture, $\kappa_{\rm 1mm} = 0.0114$ cm$^{2}$ g$^{-1}$ is the dust
opacity per gram of gas at 1.1 mm  and $B_{\rm 1mm}(T_D)$ is the
Planck function evaluated at the dust temperature \citep{enoch09}.
Applying this criterion to estimate the envelope mass for each source,
we find that {\it all} the sources have envelopes exceeding 0.1 $\msun$
so that their evolutionary stage would be correctly identified as embedded protostars. 
The table illustrates the troubling problem of associating classes
with physical stages. Some sources that are apparently Class I, and
hence inferred to be older than the Class 0 lifetime of $\sim 0.1$ Myr, are in fact
much younger. Since the length of the Class 0 and
Class I lifetimes are calculated statistically, 
for a large number of sources the effect of viewing angle will be minimized.
However, the SEDs of individual sources may be
misleading, and a fair fraction ($\sim 10$\%) of observed sources are inferred to
be older than they actually are. 

\begin{deluxetable*}{lccccc}
\tablewidth{0pt}
\tablecolumns{6}
\tablecaption{Number of protostars in each category if the angle views
  are independent observations.
\label{tabletbol}}
\tablehead{
Time & Early Class 0 & Late Class 0 & Early Class I & Late
  Class I & Class II \\
\colhead{}& \colhead{$T_{\rm bol}\leq 50$ K\tablenotemark{a}} & \colhead{ 50 K$<T_{\rm bol}\leq$100 K}  &  \colhead{100 K$<T_{\rm bol}\leq$300 K} &  \colhead{300 K$<T_{\rm bol}\leq$650 K} &  \colhead{650 K$<T_{\rm bol}\leq$2800 K} } 
\startdata
~0 kyr & 20 & & & & \\
15 kyr & 59 &1 & & & \\
30 kyr & 53 &4  &2  &1 & \\
60 kyr & 233 & 59 &22 &3 &3 \\
\enddata
 \end{deluxetable*}

\section{Comparison with An Analytic Model Grid}

In this section, we compare the known simulated source properties with those
inferred from fitting the R06 models to the SEDs.
Rather than using the full spectral information, we interpolate the
fluxes at commonly observed wavelengths. We include wavelengths 1.25,
1.65, 2.17 $\mu$m to represent Two Micron All Sky Survey (2MASS)
data, 3.6, 4.5, 5.8, 8.0  $\mu$m to represent Spitzer IRAC, 24,
70, 160 $\mu$m to represent Spitzer MIPS, and 1.1 mm representing data
from the Bolocam continuum survey at the Caltech Sub-millimeter
Observatory (CSO). 

The R06 models assume a dust extinction ratio of
$R_v=3.6$, which is slightly higher than what we adopt in our
simulation and radiative post-processing. However, the R06 adopt 
different dust models for the disk than for the envelope and outflow.
For the densest parts of the disks, R06 adopt the ``disk midplane'' model described by
\citet{whitney03a}. This model has been shown to be a good fit for the
SEDs of protoplanetary disks such as HH 30 \citep{wood02}. For the remainder of
the dust, R06 adopt the \citet{kim94} model,
which has an average particle size only slightly larger than dust in
the diffuse ISM. Neither model includes icy grain coatings, which are
expected to increase the optical depth by a factor of $\sim$ 2 \citep{chakra05}. The
former disk midplane model is more similar to the \citet{semenov03} model we
use in ORION and in the HYPERION post-processing. The
\citet{semenov03} model predicts an opacity 7.4 times larger at 1 mm and $
\sim 2$ times larger at 0.1 mm than the R06 dust model. We discuss
further in section \ref{menv} how opacity difference affects our results.

The R06 models include luminosity
from the central
star \citep{siess00} and luminosity due to accretion from the disk
onto the protostar:
\begin{equation}
L_{\rm acc} = f_{\rm acc} \frac{G M_* \dot M_*}{R_*}.
\end{equation}
R06 adopts $f_{\rm acc}=1$, which assumes that all the accretion
energy is radiated away. In contrast, ORION adopts $f_{\rm acc}=0.75$
to take into account that accretion may be radiatively inefficient and
some energy may be advected into the star or  
drive a mass outflow \citep{ostriker95}.

\begin{figure}
\epsscale{2.25}
\plottwo{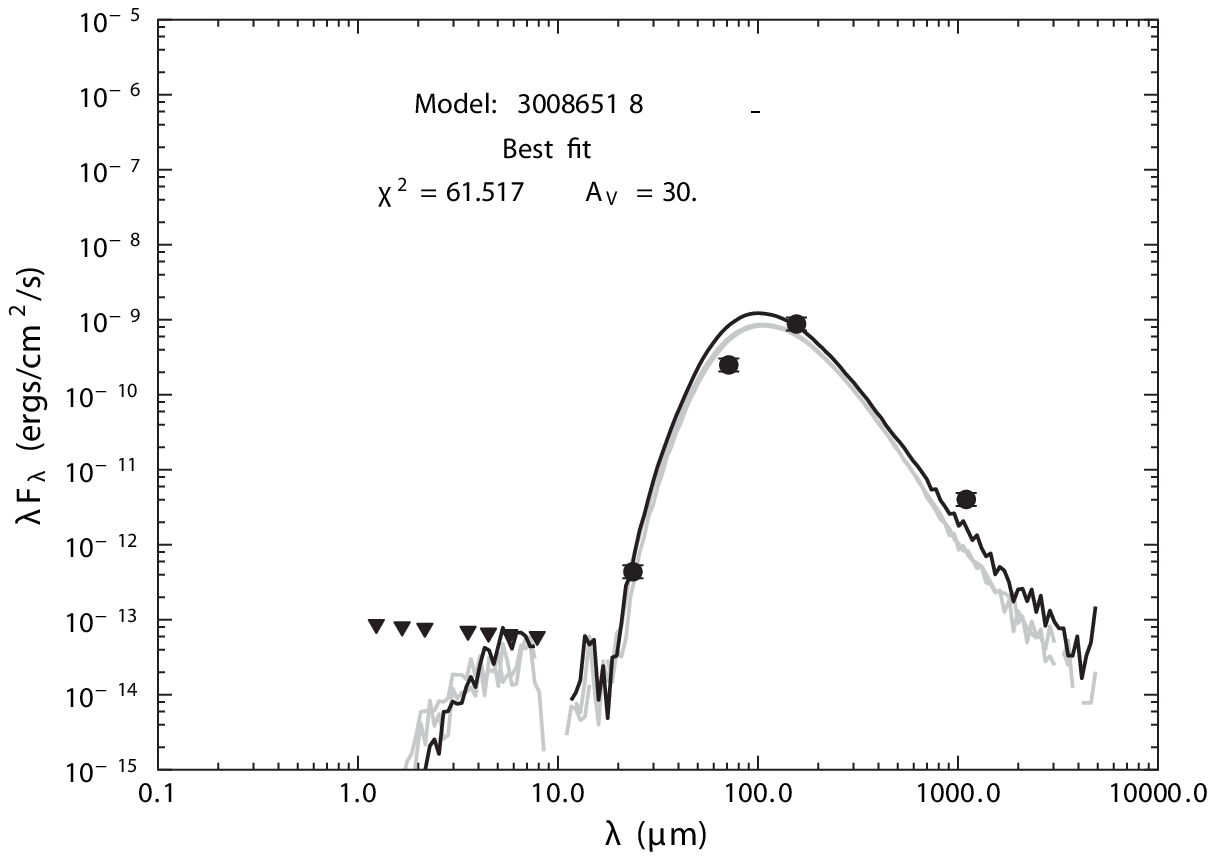}{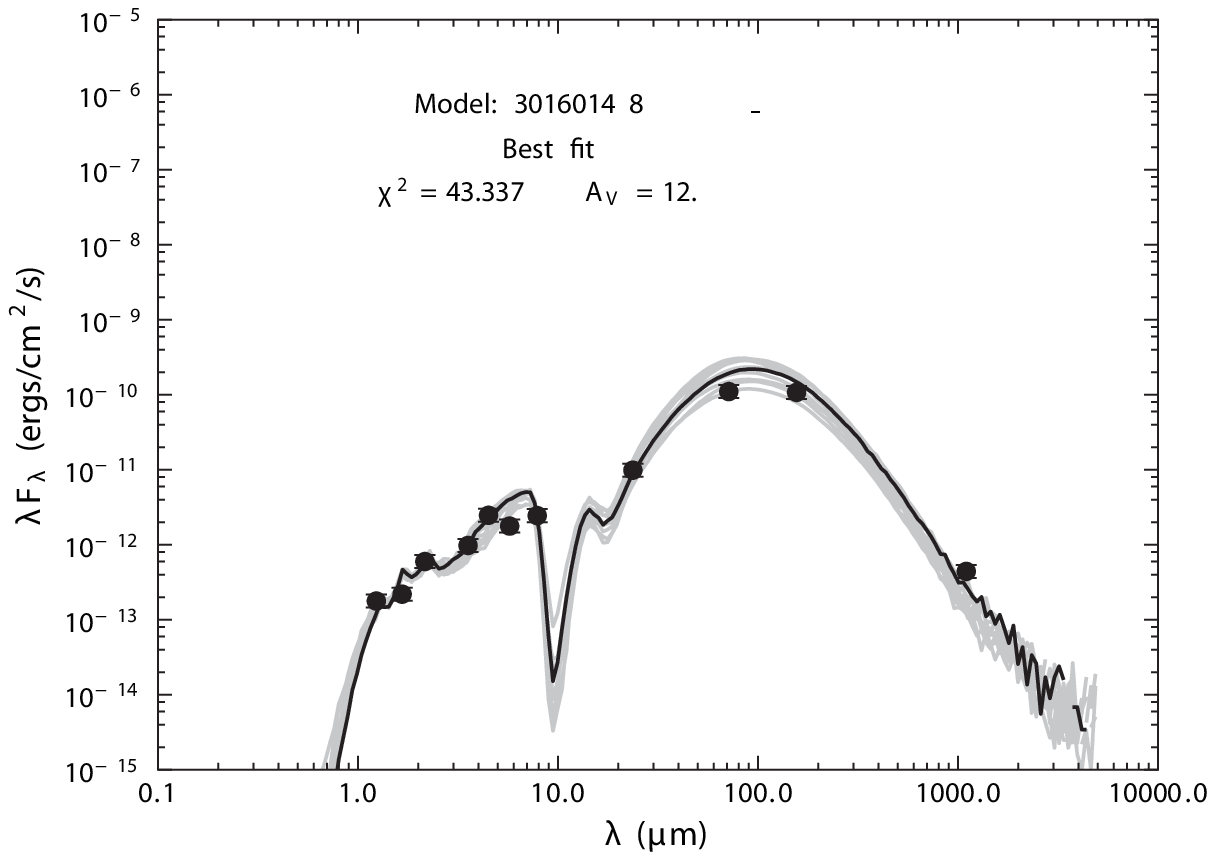}
\caption{Observed fluxes in an 1000 AU aperture for sources 1c (top) and 2c (bottom). Black points indicate the
  flux in the 2MASS, IRAC, MIPS, and BOLOCAM bands, where the triangles
  indicate upper limits imposed by the sensitivity of the instruments.
  The black line indicates the best-fit model, while the gray lines
  illustrate models with good $\chi^2$. The best $\chi^2$,
  dust extinction, scale, and best-fit model number are shown at the top. 
\label{robitseds} }
\end{figure}

In the following model comparison, we consider only those sources with
``good'' 
fits, which we define as $\chi^2 - \chi_{\rm best}^2 < 3
N_{\rm data}$, where $N_{\rm data}$ is the number of flux data points
considered. The model with the
lowest $\chi^2$ given by  $\chi^2 < 30
N_{\rm data}$ is the ``best-fit'' model. In a few cases there is only
a single model that satisfies these criteria, but in most cases
there are a number of good-fit models.
If a second source is within 1000 AU, we only
analyze the model parameters for the primary.
Figure \ref{robitseds} shows two sources with the corresponding best-fit models
overlaid. The younger source (1c) has only upper limits for the 2MASS
and IRAC bands. In general, we find that the models do not produce
good fits for the most deeply embedded sources (e.g., all sources
observed at 0 kyr and 15 kyr). This is partially due to 
``non-detections'' in the IR, but also due to the source 
faintness. Two specific aspects of the R06 models make them
unsuitable for fitting very young sources. First, the minimum model stellar
mass is 0.1 $\msun$. In this case, the poor-fits are reassuring since
the most massive star at 15 kyr is 0.05 $\msun$. The second aspect is
that the dust temperatures in the R06 models do not fall
below 30 K. Since the dimmest sources are surrounded by 10-20 K dust, they will naturally be colder than
expected by the models. 

Figures \ref{compare1}, \ref{compare2}, and \ref{radius} show the ratio of various
model properties to the actual simulation values. Overall,
the R06 models utilize more than 20 parameters (14 of which are independent)
describing the protostellar source, disk, outflow, and envelope. Here,
we restrict ourselves to comparing to eight fundamental parameters: stellar mass, stellar radius, stellar accretion rate, source
inclination, envelope mass, disk mass, outer disk radius, and inner
disk radius. The root mean square of the difference between the models and
actual value are given in Table \ref{tablerms}. We leave a more detailed comparison to future work.
Also note that we compare each source for a single view (arbitrarily oriented
with respect to the disk plane) and even
though the source inclination is considered in the $\chi^2$ fit, the
accuracy of the inferred parameters may depend upon the source orientation.

\subsection{Evolutionary Stage} \label{evolstage}

The primary physical quantity that can be derived from an SED
  is the evolutionary stage of the source. Since the R06 models cover
  a wide range of evolutionary stages, from embedded protostars to
  pre-main-sequence stars surrounded solely by low-mass disks, we 
  first examine whether the models correctly identify that all the
  sources in the simulation are in the embedded protostellar
  evolutionary stage.

R06 defined three main stages of evolution based on the value
  of the stellar mass, $M_*$, and the envelope infall rate,
  $\dot{M}_{\rm env}$. The latter is in fact a direct proxy for
  the envelope density, which is defined according to the rotationally flattened
  infalling envelope models of \citet{ulrich76}. In this classification scheme, Stage I sources (or envelope-dominated sources) have
$\dot{M}_{\rm env}/M_*\ge10^{-6}$\,yr$^{-1}$, Stage II sources
(disk-dominated sources with low-density or no circumstellar
envelopes) have $\dot{M}_{\rm env}/M_*<10^{-6}$\,yr$^{-1}$ but $\dot M_{
  *}/M_*\ge10^{-6}$\,yr$^{-1}$, and Stage III sources (sources with
little circumstellar material) have $\dot{M}_{\rm
  env}/M_*<10^{-6}$\,yr$^{-1}$ and $\dot M_{*}/M_*<10^{-6}$\,yr$^{-1}$. 

We find that all synthetic sources well
fit by the R06 models have good-fit models with $\dot{M}_{\rm
  env}/M_*>6\times10^{-5}$\,yr$^{-1}$. Thus, the R06 models clearly
indicate that the simulated sources are embedded Stage I sources. In
other words, none of the models providing a good fit are disk-only
models viewed at an extreme viewing angle.

Note that for the purposes of the model fitting
there is no direct luminosity associated with $\dot M_{\rm env}$;
instead models with high values of $\dot M_{\rm env}$ indicate the
presence of a very dense
envelope. The accretion rate associated with the output
luminosity is $\dot M_*$, the rate at which mass is dumped from the
disk onto the source  (see comparison in section \ref{accrate}). 
For a realistic protostellar evolutionary sequence, there is likely a strong correlation between the
infall rate and the stellar accretion rate. However, in the R06 model
space these two rates are independent.

Physically, the infall rate may be
many times higher than the stellar accretion rate if there is mass removal due to protostellar
outflows and multiplicity (such that the infalling gas is distributed
among two or more protostars). For example, \citet{hansen12} finds a core
efficiency factor of $1/3$, which means that $2/3$ of the infalling
core gas is expelled by outflows and thus $\dot M_*$ is correspondingly lower.
In \citet{Offner10}, infall rates in radiation-hydrodynamic
simulations without outflows fall in the range of $\sim 10^{-6}~\msun
{\rm yr}^{-1}-6\times 10^{-5}~\msun {\rm yr}^{-1}$. These are at least
a factor of 2-20 times lower than the infall rates suggested by the R06 best-fit
models, which predict $\sim 0.2-3\times 10^{-4} \msun{\rm yr}^{-1}$. If the cores are turbulent, as
they are here, and
not simply undergoing freefall collapse then the 
infall rate may be expected to be a factor of $\sim 2$ lower than suggested by the 
\citet{ulrich76} model \citep{mckee03}. Given these very high predicted infall
rates, the stellar accretion rate {\it must} be significantly lower in
order to avoid over-estimating protostellar luminosities. We will show
that this is the case in section \ref{accrate}.

\subsection{Stellar Mass} \label{stellarmass}

\begin{figure}
\epsscale{1.20}
\plotone{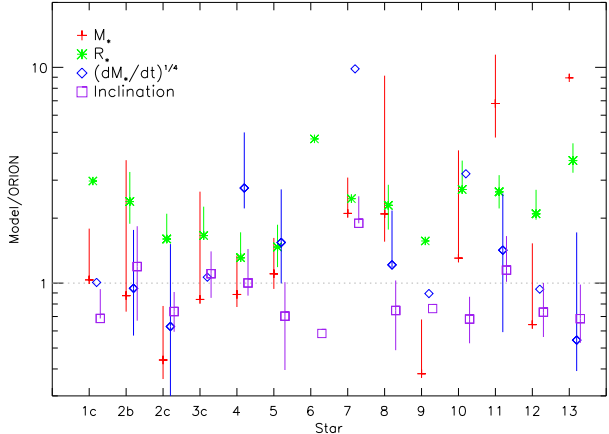}
\caption{Ratio of the inferred best model values to the actual
  simulation value for each of the sources with good fits, where $M_*$ is the stellar mass, $R_*$ is the
  stellar radius, $(dM_*/dt)^{1/4}$ is the fourth root of the accretion
  rate onto the star, and the Inclination refers to the tilt with
  respect to the line of sight. The dotted line indicates where the models determine a value identical to the true value in the simulations.
 Points with no error bars have
  only one best fitting model parameter. 
\label{compare1} }
\end{figure}

Figure \ref{compare1} illustrates that inferred masses tend to be more
accurate for the older sources, where the true masses lie within or
very near the model range (e.g., 1c-6). The masses tend to be
overestimated for the younger objects, which are somewhat contaminated by 
dust heated by neighboring protostars. Since the R06 models only include models with
stellar masses $\ge 0.1~\msun$, they will tend to
over-estimate the masses of sources near this limit or below by construction. 
Overall, R06 best-fit models achieve the best accuracy for bright
sources in 
non-clustered regions. However, only three of the 13 sources have best-fit models that disagree by more
than a factor of 2. Given that the models span two orders of
magnitude in source mass, the agreement between simulation and model
is reasonable.

\subsection{Stellar Radius}

In contrast to the stellar mass, the R06 models
systematically overestimate the stellar radius, generally by a factor
of $\sim$2-3. This is understandable since the R06 models use stellar properties interpolated from the non-accreting isochrones of
\citet{siess00}. In contrast to \citet{palla91,palla99}, who agree well with
the evolutionary model we adopt in ORION, the \citet{siess00}
models do not include accretion. Instead, all protostars are
initialized with their final stellar mass and a stellar structure determined by
hydrostatic equilibrium for central temperatures $<10^6$ K, i.e.,
pre-deuterium burning. Consequently, the \citet{siess00} protostars begin with
inflated radii $>4 ~\rsun$ and contract towards the zero age main
sequence.\footnote{In the context of the R06 models, the ``initial'' radius
  is given at 10$^3$ years, the earliest source
  age included in the models.}  For \citet{palla99}, protostars with accretion rates of
$10^{-5} ~ \msun$ yr$^{-1}$ terminate their accretion phase after 0.1
Myr on an $L-T$ line in the H-R diagram referred to as the stellar
birth-line. This birth-line coincides fairly closely 
with the \citet{siess00} H-R placement of stars with ages of 0.1 Myr and
masses $M<1\msun$ (e.g., \citealt{dario10}). Consequently, the two sets of models give similar values
for the stellar radii as the stars approach 0.1 Myr. For stars with
ages $\lesssim 0.05$ Myr, the \citet{siess00} models overestimate the
radii relative to our ORION model by a factor of $\sim$2. \citet{palla91}
find that assuming lower accretion rates, e.g. $10^{-6} \msun$ yr$^{-1}$,
produces smaller pre-birth-line radii. Thus, for protostars with low
accretion rates and young ages, the two models will be even more
discrepant. 
Some of the difference in model radii is offset by the ORION
assumption that the accretion energy is not perfectly radiated
away, i.e., $f_{\rm acc}=0.75$. This results in an effective
stellar radius of $(1/f_{\rm acc})R_*$, which is somewhat closer to the
estimated values of \citet{siess00}

Despite clear model differences, determining the absolute correctness of the stellar evolutionary
models is not straight-forward.
The initial protostellar radius, its subsequent evolution, and how
it depends upon accretion and radiative efficiency, remains hotly
debated \citep{hosokawa11, baraffe09}. Even without considering these
effects, various calculations often adopt initial radii of 2.0$\rsun$ \citep{stahler88} to 3.5$\rsun$
\citep{palla91}. In cases where
accretion is considered, accretion rates comparable to
$\sim10^{-6}\msun$yr$^{-1}$, which are appropriate for low-mass
stars, tend to have early radii of $\sim 1.5 R_{\odot}$
\citep{palla91, tan04, hosokawa09}. 
Given the differences between stellar models, it is encouraging
that many of the protostellar masses inferred are nonetheless quite close to the
simulated values (see section \ref{stellarmass}).
\begin{deluxetable}{lcc}
\tablewidth{0pt}
\tablecolumns{3}
\tablecaption{Root Mean Square difference between the actual ORION and
  R06 best-fit values for each parameter and the median
  ratio of the best-fit R06 and ORION values.
\label{tablerms}}
\tablehead{
Parameter\tablenotemark{a} & RMS\tablenotemark{b} & Median$({\rm R06/ORION})$ } 
\startdata
$M_*$ & 0.4 $\msun$& 1.1\\
$R_*$ & 2.8 $\rsun$& 2.4\\
$dM_*/dt$ & $6.8\times 10^{-7}$ $\msun$ yr$^{-1}$& 2.2\\
Inclination  &  18.5 deg & 0.7\\
$M_{\rm env}$ & 0.9 $\msun$& 3.2\\
$M_{\rm env,4840AU}$ & 57.0 $\msun$& 6.2\\
$M_{d}$  & 0.02 $\msun$& 11.2\\
$R_{d}$  & 174.1 AU & 0.3\\
$R_{d,in}$  & 27.6 AU &  0.3\\
\enddata
\tablenotetext{a}{All parameters are compared assuming a 1000 AU
  aperture except $M_{\rm env,4840AU}$, which is calculated using a 4840
AU aperture.}
\tablenotetext{b}{The RMS is defined as ${\rm RMS} = \left(\sum\limits_{i=1}^N
  \frac{\left[O_i-R06_i\right]^2}{N} \right)^{1/2}$.}
\end{deluxetable}

\begin{figure}
\epsscale{1.20}
\plotone{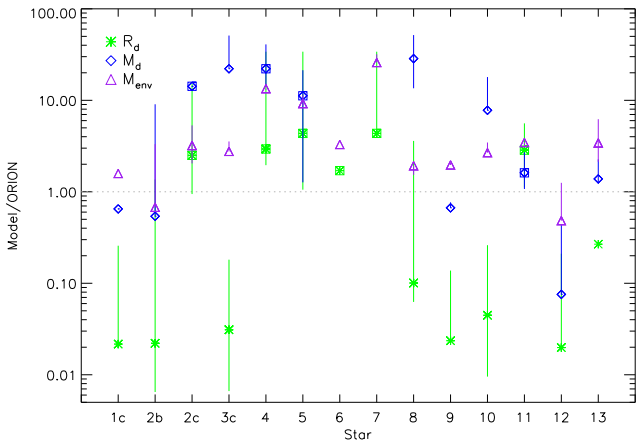}
\caption{Ratio of the inferred best model values to the actual
  simulation value for each of the sources with good fits, where $M_d$ is the disk mass, $R_d$ is the
  stellar radius, and $M_{env}$ is the envelope mass. The dotted line indicates where the
  simulation and models are equivalent. Points with no error bars have
  only one best fitting model parameter. Boxed points 
  indicate that the comparisons use the simulation upper limits for
  the disk mass or radius.
\label{compare2} }
\end{figure}

\subsection{Stellar Accretion Rate} \label{accrate}

The R06 models achieve fairly good agreement with
$\dot M_*$. With the exception of two cases, the models either bracket the actual value or
come very close to it. 
Since the good-fit models may span four orders of
  magnitude in the accretion rate, we plot the fourth root to reduce the error bar extent in
  the plot.

In the R06 models, the range of model accretion rates decreases
monotonically with source age. If protostars experience episodic
accretion \citep{hartmann96, zhu09, vorobyov10, Offner11}, then
protostellar accretion rates may jump between $10^{-7} \msun$
yr$^{-1}$ and  $10^{-4} \msun$
yr$^{-1}$, extremes which may not be spanned by the models and thus result
in age or mass errors. However, we find that, at least for these early times,
variability in the instantaneous accretion
rate is less than an order of magnitude. Consequently, our simulated values
fall within the model ranges of R06.  The extreme accretion events
known as FU Ori outbursts are believed to result from a combination of
thermal and magnetic instability in the inner accretion disk \citep{zhu09}. We
do not treat these effects in these simulations, and thus all accretion
variability is due to gravitational instability within the disk or turbulent
variations of the core envelope.   

The good agreement between the stellar accretion rate determined from
  the SED models and the accretion rate in the simulation is
  surprising, nonetheless, since the HYPERION source modeling assumes a
  single input spectrum with a bolometric luminosity equal to the sum of the
  stellar and accretion luminosities. Note that this sum is ultimately what
  sets the range of possible good-fit values for the protostellar masses and
  accretion rates shown in Figure \ref{compare1}.
  However, since there are many different values for the stellar and
  accretion luminosities that give the same bolometric luminosity,  it should not be possible to
  disentangle the relative contributions from these from SED
  modeling.  Instead, it is likely that the accretion rate is only
  constrained compared to the original range in the model grid due to
  other parameter constraints. In particular, due to the parameter space sampling in the R06
  grid, models with higher envelope infall rates have higher
  stellar accretion rates, and conversely, models with lower
  envelope infall rates have lower stellar accretion rates (see
  Figure 1 of \citealt{robit08}). Thus,
  constraining the envelope infall rate (or density), automatically
  restricts the stellar accretion rate to a range that is
  smaller than that of the whole set of models. Similarly,
  constraining the total luminosity also automatically rules out
  models with stellar accretion rates that are too high.

\begin{figure}
\epsscale{1.10}
\plotone{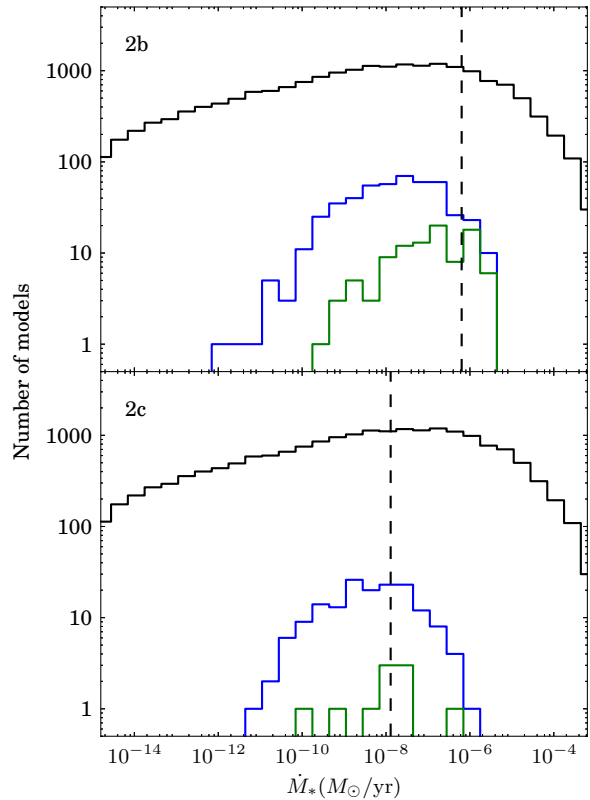}
\caption{
The top and bottom panels show, for two different sources, the
distribution of stellar accretion rates, $\dot M_*$, in the
entire R06 set of models (black), the models that
provide a good fit to the SED (green), and the models in R06 that have
envelope infall rates and bolometric luminosities within $\pm$10\% of
the values for each good fit. The vertical dashed lines indicate the
best-fit model.
\label{mdotdisk}}
\end{figure}

Figure \ref{mdotdisk} shows the distribution of stellar
accretion rates for all models in the R06 set, for the models providing
a good fit, and for all the models in R06 that have envelope infall
rates and bolometric luminosities within $\pm$10\% of the values of
the good fits.\footnote{Since the infall rates and bolometric luminosites have
been varied \textit{independently} of the other fit parameters, these models are not
necessarily close, i.e., within 10\% of being good-fit models, to the data.} The figure demonstrates that while the range of accretion rates
for the good fits is much narrower than the range in the entire set of
models, this range is primarily set by constraints on other
parameters. It is therefore likely that $\dot M_*$ only
appears to be well fit due to constraints on other parameters that can
be more directly determined, such as the envelope infall rate and the
bolometric luminosity. These two parameters supply the main
  constraints on the stellar accretion rate either by setting an upper bound
(the luminosity) or by restricting the range for
  given envelope properties (the infall rate). However, we note that the agreement does appear
to indicate that the accretion rates assumed in the R06
models span a realistic range.

\subsection{Inclination}\label{inclination}

Inferring the correct source inclination is important for accurately
determining other source properties. If a source is observed
edge-on, but not identified by the models at this orientation, then it
will appear erroneously younger. We use the simulated angular momentum of the
protostar to determine the inclination with respect to the line of
sight, where the net angular momentum vector is assumed to point perpendicular to
the disk plane. Figure \ref{compare1} illustrates that the most
discrepancy occurs where the inclination is misfit, e.g.,
stars 8 and 9. These sources are inferred to have higher stellar
masses and lower
accretion rates than they actually do. At intermediate inclinations between pole-on
and edge-on the inclination parameter is fairly degenerate. As a result, the
inclination constraints are ultimately not very tight.

\subsection{Envelope Mass}\label{menv}

For each R06 model, the envelope radius, $R_{\rm env}$, is randomly
drawn from a uniform logarithmic distribution of values between
$R_0 \times 4$ and $R_0/4$, where $R_0$ is approximately the radius at
which the optically thin radiative equilibrium temperature declines
to 30 K:
\beq
   R_0 = \frac{1}{2} R_* \left(\frac{T_*}{30 \rm{K}} \right)^{2.5}.
\eeq
These values are further constrained to lie between $10^3$ AU and
$10^5$ AU. Inside this radius, the density profile goes as $\rho(r)
   \propto r^{-3/2}$ with some slight modification due to rotational
   flattening. The envelope mass is then computed by integrating the mass inside 
$R_{\rm env}$. The ambient gas outside the envelope is assumed to be a
   constant density randomly drawn from the range 50 cm$^{-3}\leq
   n_{\rm H_2}/(M_*/\msun) \leq $ 100 cm$^{-3}$. Thus, the envelope
   mass is coupled to both the density distribution and stellar properties.

In the simulations, however, the gas morphology is independent of the
stellar properties and is not nicely spherically
distributed. The mass asymmetry means that there is no clear cut way to define a physically meaningful
envelope mass. Instead, we follow the R06 model convention and
calculate the total mass inside a spherical volume with the radius
$R_{\rm env}$ suggested by the best fit model. This circumvents the need to define
a density minimum or an effective radius for the simulations.

\begin{figure}
\epsscale{1.20}
\plotone{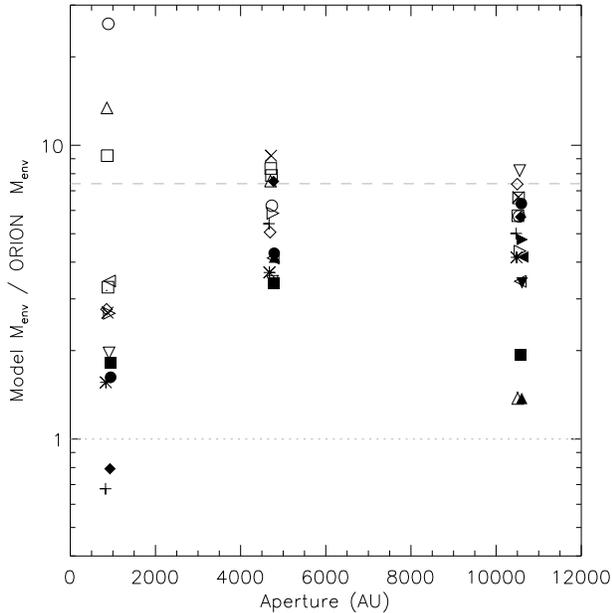}
\caption{Ratio of the inferred best-fit envelope mass to the actual
  envelope mass for each of the sources with good fits as a
  function of aperture size. The dotted line indicates where the
  simulation and models are equivalent. The dashed line indicates the
  ratio of the ORION opacity to the R06 model opacity (7.4) at 1
  mm. All inferred envelope masses are shown (not only those centered on
  the most massive protostar in a multiple system). Each source has a
  unique symbol.
\label{menvap} }
\end{figure}

Figure \ref{compare2} shows that the model envelope masses run
high relative to the simulated envelope masses. In very few cases do
the best fit models encompass the true value. This is primarily a result
of the different opacity models adopted by ORION and R06. 
The envelope mass is determined from the long wavelength
emission. At 1mm, the \citet{semenov03} opacity is 7.4 times higher than
R06 dust opacity; it is
$\sim 2$ times higher at 0.1 mm. Since the envelope mass is inversely
proportional to the opacity (see eq.~\ref{menv}), a given flux will be interpreted by the R06 models to correspond to a
larger mass envelope. Figure \ref{menvap} shows that the discrepancy
is less for a 1000 AU aperture than for the larger apertures,
which include more of the cold, extended envelope. 

However, in all cases there is a fairly large scatter in
the best-fit models. Since the simulated cores are embedded within a
cloud and somewhat clustered,  it is likely
that some of the scatter is due to difficulty separating envelope
emission from foreground emission, and thus correctly determining the
line-of-sight extinction. The value of the inferred envelope radius is
also important since most of the envelope mass is found a large radii.
Since the R06 models only include envelope dust with
temperatures $\gtrsim 30$ K, they preferentially neglect the coldest
envelope mass. This can contribute to error in the 1 mm flux data
point, which influences the determination of the envelope mass.
Despite this innaccurary, the models do correctly
identify the SEDs as belonging to embedded protostars and do not mistake them
for more evolved disk-dominated sources.

For a given observed SED there will always be some underlying uncertainty in
the grain composition and size distribution, which depends upon the
local temperature, metallicity, and source age. However, our
comparison suggests that without icy coatings and inclusion of cold
envelope gas, the R06 models may systematically overestimate observed
envelope masses.



\subsection{Disk Mass}

We follow the same procedure as \citet{Offner10} to define what
constitutes a disk in the simulations. First, we identify all connected
cells around the protostar with densities $\ge 10^{-15}$ g
cm$^{-3}$. This threshold effectively selects for gas comprising the
disk, since the density of the gas in the envelope falls off rapidly.  We then estimate the total angular momentum vector of the
gas and rotate the coordinate frame so that the net angular momentum
vector is parallel to the z-axis. Finally, we restrict the vertical
disk height in the z direction to $\pm$5 cells from the disk
mid-plane. This is to eliminate confusion with streams of gas feeding
the disk. The disk mass is then the integrated gas mass in these cells.

Due to the inclusion of outflows, we find that
the disks are less orderly than in the simulations
analyzed in \citet{Offner10}.
In a few cases, i.e., where the protostellar mass is $\lesssim
0.1~\msun$, there is no gas exceeding the density threshold. For
these, we assume that the disk is too small to be resolved by the
simulation and set the disk radius, $R_d$, to 4 fine cells (16 AU), the radius of the
outflow launching region, and the disk mass to the mass enclosed in a
spherical volume with this radius. These values thus represent upper limits on the
disk mass and radius.
 
Figure \ref{compare2} shows that the model disk masses are not well
correlated with the simulated disk masses. The models tend to
overestimate the disk mass, and in some cases the
discrepancy is a factor of 10. Unlike other parameters, such as the
stellar mass, the actual disk mass does not fall within the range of
good-fit models in nearly all cases. This is mainly because the
simulated disks are more complex and 
less dense at the mid-plane, where the grid resolution does not resolve
the scale height. The mass estimates are most overestimated in cases where the
R06 model underestimate the size of the inner hole (see section
\ref{innerdisk}), and thus overpredict the mass in the inner disk
region. 
Since the simulated disks are more extended and less
dense, it is also challenging for the models to accurately distinguish
between emission from the disk and emission from the envelope, which
contributes to the scatter in the agreement.

\subsection{Disk Radius}

In the R06 models, the disks are assumed to be perfectly symmetric with a sharp edge, so
there is no difficulty in defining the disk radius. In the
simulations, the disks are often asymmetric and occasionally have
spiral structure that winds out to larger distances. For the
simulations, we define the radius as the average radial distance of
all the cells in the disk. This minimizes the effect of asymmetry on
the radius estimate. 

Figure \ref{compare2} shows that the model disk radii are
generally much smaller than the simulated disk radii. 
The best-fit model disks have a
median radius of $\sim$32 AU with a minimum value of 5.7 AU, while the simulations have a median of
$\sim$136 AU and minimum (upper limit) of 16 AU. 
This is to be
expected since the simulations do not resolve the disk scale height
and thus likely over estimate the disk extent. Since the disk mass is
distributed over a larger volume and the gas is less dense than expected by the
R06 models, some of the gas likely resembles envelope
gas for the purpose of model fitting. However, the large range of
radii inferred by the good-fit models suggests that this 
parameter is not well constrained by the SED in general. 
The sources where the model agrees within a factor of two are all comparisons assuming the upper limit of
16 AU for the simulated disk radius.
In principle, the
disk properties should be sensitive to the viewing angle, however, we
don't find that sources with better fit inclinations have necessarily
more accurate disk properties. The accuracy of the inferred
  disk properties also likely depends on evolutionary stage. For 
  more evolved sources, where the envelope mass is small, the R06 may
  determine more accurate disk masses and radii.

\subsection{Disk Inner Radius}\label{innerdisk}

We choose to ignore the gas inside the accretion region, which
corresponds to the $8^3$ cells centered on each star. 
This effectively
enforces that all accretion disks have an inner radius, or hole, of
$\sim$ 16 AU.
Since gas is removed
from these inner cells during the accretion process at each timestep the dynamics in these cells is
predominantly  
numerical. The outflow is launched just outside these cells so that gas
remains close to the star when otherwise it might be evacuated by the
outflow. To account for this, we set the density to zero in these
cells when observing the protostars with HYPERION. 

\begin{figure}
\epsscale{1.20}
\plotone{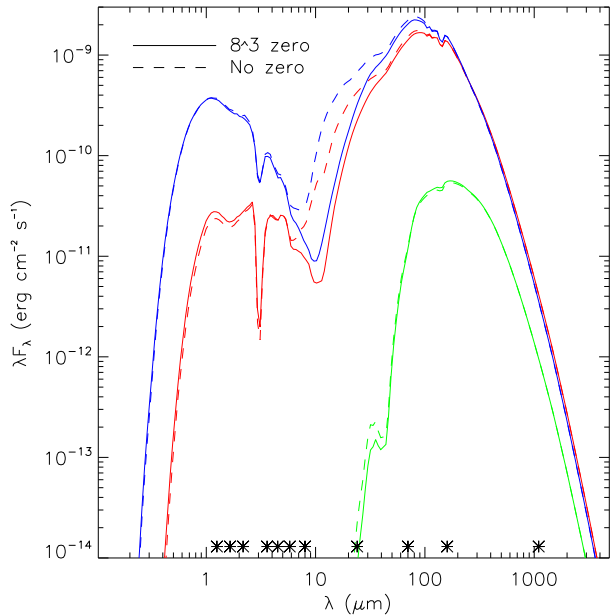}
\caption{Three SEDs calculated with (dashed) and without (solid) the
  gas inside the accretion region. The three classifications are early
  Class 0 (green), late Class 0 (red) and late Class I (blue).
\label{zero} }
\end{figure}

For Class 0 sources, which are heavily embedded, the details of the
inner region have minimal effect 
on the SED characteristics. Figure \ref{zero} shows three
SEDs calculated with and without the accretion region gas. Including the
accretion gas produces an increased flux from $\sim 8-20 ~\mu$m, i.e., increased
photon scattering off of the inner warm dust. In the early Class 0 case
(e.g., the 
green curve in Figure \ref{zero}), the dust is optically thick to 
$<20\mu$m emission and the inner region has
negligible influence. The SEDs of late Class 0 and I sources
are more sensitive to this emission. In Figure \ref{zero} the
bolometric temperatures change by 3.6\% (late Class 0) and 7.8\% (late
Class
I), while the early Class 0 bolometric temperature only increases by 0.2\%.
However, the difference turns out to have little effect on the
properties inferred in the model comparison, which uses only discrete
wavelength points from the SED. The largest flux difference falls {\it
  between} the observed 8$\mu$m and 24$\mu$m data points. 
Consequently, the inferred model properties
are largely insensitive to the gas in the accretion region. (See
Appendix A for additional discussion.)

\begin{figure}
\epsscale{1.20}
\plotone{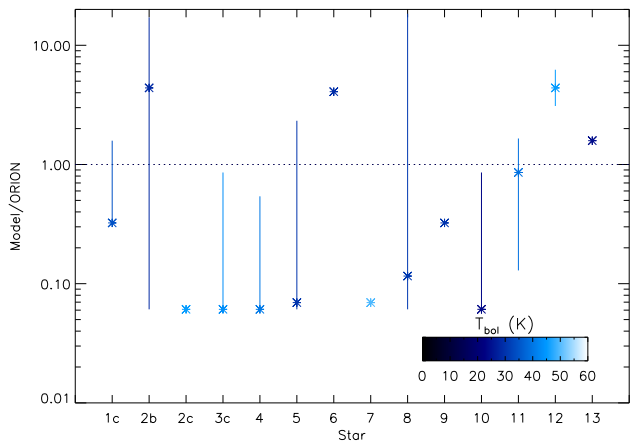}
\caption{Ratio of the inferred best model inner disk radius to the actual
  simulation value for each of the sources with good fits. The dotted line indicates where the
  models determine a value identical to the true value in the simulations. Points with no error bars have
  only one best fitting model parameter. The color indicates the
  bolometric temperature.
\label{radius} }
\end{figure}

We can verify that this is the case by comparing the inner disk
radius inferred by the best fit models with the actual inner disk
radius of 16 AU. The R06 models include disks with inner holes as large as 100 AU in order to account for the
possibility of a close, unresolved stellar companion that is evacuating the
inner disk. Figure \ref{radius} shows the ratio of the
inferred and actual inner disk radius. The large range of good-fit 
radii and large discrepancies of the best fit model suggest that the SED shape
is not particularly sensitive to the inner hole size. At least for Class 0
objects, the accuracy is also insensitive to the magnitude of
the bolometric temperature. 

\section{Conclusions}

We present results from turbulent, gravito-radiation-hydrodynamics AMR
simulations of forming stars. By including the effects of protostellar outflows and
radiative heating, we are able to follow the evolution of the
protostars self-consistently using realistic physics. We neglect
magnetic fields, which we leave to be addressed in future work.

We use the Monte-Carlo radiative
transfer code HYPERION to compute the synthetic spectral energy distributions of
the sources for a variety of apertures, viewing angles, and
resolutions. We find that the HYPERION and ORION calculated dust
temperatures are most discrepant in shock heated gas such as within
the outflow cavity, but generally agree within a factor of two for most cells.

Synthetic images of the forming protostars in the infrared and
(sub) millimeter show clearly recognizable observational features.
Emission from $\sim$1-30 $\mu$m traces out the outflow cavity. In the
sub-millimeter, an edge-on dense disk appears as a dark band obscuring the protostar.

Like previous work, we find that the source SED and its
inferred properties are very sensitive to the viewing angle. The
inferred bolometric luminosity variation can exceed a factor of 5 depending
on the view. This corresponds to a typical standard deviation in log
luminosity of 0.2. However, in most cases, the inferred range includes the true luminosity. The
 variety of SEDs also corresponds to a large range of bolometric
 temperatures, which can often span two spectral
 classifications. This underscores the difficulty of associating
 classes with specific evolutionary states based on SED
 characteristics alone. Since star formation is clustered, we find
 that the later forming stars suffer environmental contamination from
 the neighbors, increasing the uncertainties in their classification
 and inferred properties.
 
We next assess the accuracy of source properties inferred from analytic
models. We use the model grid of \citet{robit06} to derive best-fit
parameters for each of the sources and discuss the accuracy for a
number of fundamental parameters.
No good fits occur for the
dimmest, most embedded objects, i.e., those with ages $<20$ kyr.
The sources well fit by the R06 models are correctly identified as
embedded protostars and not mistakenly fit with more-evolved disk-only
models. 
Using the R06 models, we thus come to the same conclusion
  as when assessing the observed envelope mass: all the protostars are embedded Stage I
  objects. Since the R06 models do not include stars with final masses below 0.1
  $\msun$, which is the critical envelope mass, there is small
  advantage to using the R06 models purely to determine the
  evolutionary stage in lieu of simply estimating the envelope mass. However, the R06 models offer
  many additional, potentially enlightening,  parameter constraints. Further work extending the
  R06 model grid to lower masses may facilitate the evolutionary
  characterization of lower mass, young sources.

The model parameters determined from the good SED fits for the older sources exhibit varying
amounts of accuracy compared to the true simulation
parameters. We find
that the stellar accretion rate spans the true value in most cases. The models are consistent with the true stellar mass for the older and/or
more isolated sources. This is partially because the R06 models have a
minimum mass of 0.1 $\msun$, and thus, will not well fit young
low-mass sources by construction. 
In several cases, where the source environment
is affected (via heating or outflow activity) by a more luminous
neighbor, the stellar mass estimations are quite discrepant. The
R06 models also use the \citet{siess00} stellar evolution models and
thus systematically overestimate the source radii at early
times. This may also
introduce error into the parameter estimates. 
The inferred source
inclination exhibits decent agreement with the actual inclination,
although this parameter is not strongly constrained by the fits.
The models systematically overestimate the envelope mass, a trend that
might be corrected in future models that include icy grains and do not exclude the coldest
envelope gas. We find that the disk mass and radius are not well fit by the
models. While we expect that including magnetic effects and
even better resolution would lead to smaller disks, ideally the
breadth of the models
should encompass the full disk parameter space. Some
numerical studies have found that
magnetic effects may completely suppress the formation of disks \citep{galli06,
  price07, hennebelle09},
which would also be challenging for the R06 models since they assume
that all young sources have a disk.

Overall, our study highlights a number of problems with inferring
source, disk, and envelope properties from observed
SEDs of protostellar sources. Consequently, the classification and inferred properties of
individual sources should be accepted with caution. Properties of isolated protostars are likely to
be more accurate than protostars in clustered regions.
As
summarized in Table 3,  properties such as the inferred stellar
mass may be inaccurate by more than a factor of two and the envelope mass
may be overestimated on average by a factor of four. The comparison is
performed assuming that the sources are unresolved, so additional information from
direct 
imaging might improve the accuracy of inferred parameters.

Some of our
comparisons underscore uncertainty in fundamental star forming
properties, such as dust properties and the dependence of stellar evolution
on accretion,
while others highlight improvements for future studies. A new and
improved model grid using HYPERION will aim to improve some of these
deficits (Robitaille et al.~in prep). 
Although non-magnetized
simulations show good agreement with observations \citep{bate09,
  hansen12} and we find that the resolution of the inner region has a
small affect on our results, future numerical studies
including magnetic fields and sub-AU resolution will be advantageous.


\smallskip
\acknowledgements{The authors thank Andrew Cunningham for helpful
  discussions of protostellar outflows. This research has
been supported by the NSF through grants AST-0901055 (SSRO) and
AST-0908553 (CFM and RIK), by NASA
through the Spitzer Space Telescope Fellowship Program (TPR) and ATFP
grant NNX09AK31G (RIK, CFM), and by the US Department of Energy at LLNL under contract DE-AC52-07NA (RIK).
The ORION
simulations were performed on resources managed by the National Energy Research Scientific Computing Center, which is supported by the Office of Science of the U.S. Department of Energy under Contract No. DE-AC02-05CH11231. 
The HYPERION calculations and other data analysis were performed
on the Odyssey cluster, which is supported by
the Harvard FAS Sciences Division Research Computing Group. Figure
  \ref{windvel} was rendered using yt \citep{turk11}.}

\bibliography{outflowbib.bib}
\bibliographystyle{apj}

\appendix
\section{Sensitivity of the Results to the Inner Disk Resolution}

With or without the inclusion of the accretion region gas, the inner
disk properties are not well resolved even with 4 AU cell
resolution. Observationally, the warm inner disk gas contributes SED
emission at $\le 8~\mu$m wavelengths \citep{robit06}. In observed SEDs, this emission can
be used to directly infer disk properties (e.g., disk mass, flaring, gap
size) of non-embedded protostars (e.g., \citealt{espaillat11}).  In instances where the protostar is deeply embedded,
namely during the Class 0 phase, radiation from gas close to the star is
significantly reprocessed by the cold, envelope gas. As we have demonstrated in section \ref{innerdisk}
the properties inferred by the R06 models are relatively insensitive
to the details of the inner disk in most cases.

We can estimate
the importance of the inner regions more quantitatively by estimating the 8$\mu$m optical
depth for different protostellar views. In cases where the optical
depth $\tau_8 \lesssim 1$, our models will not accurately represent the
SED at short wavelengths. For  $\tau_8 \sim$ a few, the gas is
optically thick to the emission from the inner regions, and the SEDs will not be significantly 
effected by the details of the numerical simulations on these size scales.

\begin{figure}
\epsscale{0.60}
\plotone{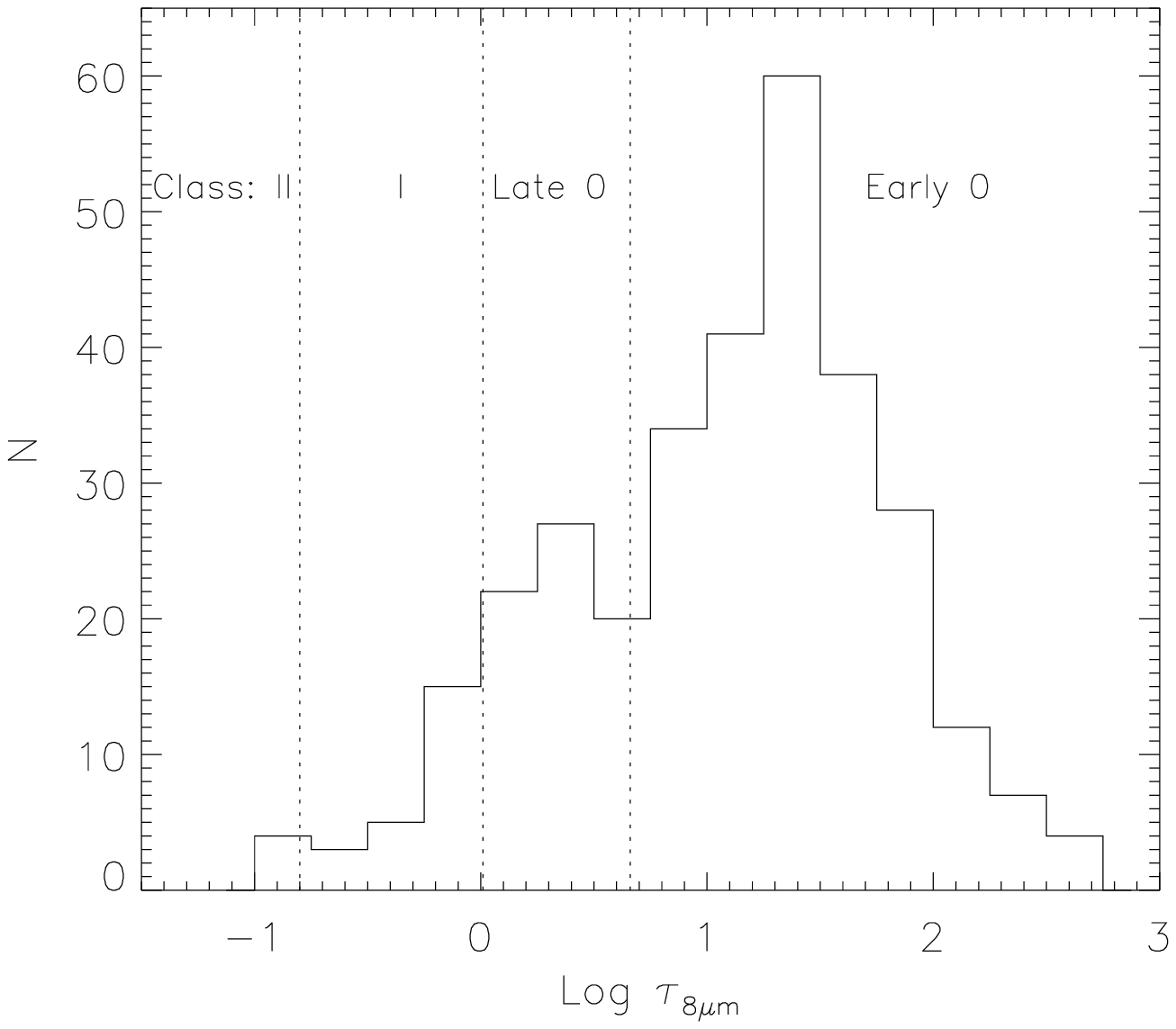}
\caption{Distribution of optical depths at 8$\mu$m for each of the
  viewing angles for the sources at 60 kyr. The vertical lines
  indicate the approximate divisions between classes (see Table \ref{tabletbol}).
\label{opac} }
\end{figure}

Figure \ref{opac} shows the distribution of optical depths at 8 $\mu$m
for all the different views of the protostars at the final
output time. Only $27$ views have optical depths $<1$. This
approximately corresponds to the number of Class I and II sources
that we find. Since these objects are viewed on sight-lines down the outflow
cavity, we would a priori expect a low optical depth.  The large
majority of sources are Class 0. Based upon these optical depths, we expect that our 
simulated early Class 0 and most late Class 0 SEDs accurately represent
{\it real} SEDs. The simulated SEDs of later classes will only accurately represent
protostellar disks with large inner holes. Our results in section 4,
which are obtained for Class 0 sources, should thus be
generally applicable and relevant to observations.

\end{document}